\documentclass{aa}

\usepackage{txfonts}
\usepackage[tbtags,fleqn]{amsmath}
\usepackage{amsfonts}
\usepackage{graphicx}
\usepackage{natbib}
\usepackage{lscape}
\usepackage{siunitx}
\usepackage{booktabs}

\usepackage[utf8]{inputenc}

\newcommand{\e}{\mathrm e}
\newcommand{\diff}{\mathrm d}

\newcommand{\R}{\mathbb R}
\newcommand{\mincir}{\raise
  -2.truept\hbox{\rlap{\hbox{$\sim$}}\raise5.truept \hbox{$<$}\ }}
\newcommand{\magcir}{\raise
  -2.truept\hbox{\rlap{\hbox{$\sim$}}\raise5.truept \hbox{$>$}\ }}


\DeclareMathOperator{\Var}{Var}
\DeclareMathOperator{\Cov}{Cov}
\DeclareMathOperator{\E}{E}
\DeclareMathOperator{\erfc}{erfc}

\newcommand{\cdbox}[1]{%
  {\color{red}%
    \dbox{\color{black}#1}}%
}
\usepackage{dashbox,framed,color,ocg-p}
\newcommand{\ToggleLayer}[2]{%
  \leavevmode
  \pdfstartlink user {
    /Subtype /Link
    /Border [0 0 0]%
    /A <<
      /S/JavaScript
      /JS (
         var aOCGs = this.getOCGs(), Layer;
         var Layers = "#1".split(","), Active = -1, i, l;
         for (l=0; l<Layers.length; l++) {
           Layer = Layers[l];
           for (i=0; aOCGs && i<aOCGs.length; i++) {
             if (aOCGs[i].state && aOCGs[i].name == Layer) {
               Active = l;
               aOCGs[i].state = false;
             }
           }
           if (Active >= 0) break;
         }
         if (Active == -1) {
           for (l=0; l<Layers.length; l++) {
             if (Layers[l] == "") Active = l;
           }
         }
         Active = Active + 1;
         if (Active == Layers.length) Active = 0;
         Layer = Layers[Active];
         for (i=0; aOCGs && i<aOCGs.length; i++) {
           if (aOCGs[i].name == Layer) aOCGs[i].state = true;
         }
      )
    >>
  }#2%
  \pdfendlink
}

\begin{document}

\title{A new method to unveil embedded stellar clusters}
\author{Marco Lombardi\inst{1}, Charles J. Lada\inst{2}, and Jo\~ao
  Alves\inst{3}} \mail{marco.lombardi@unimi.it} \institute{%
  University of Milan, Department of Physics, via Celoria 16, I-20133
  Milan, Italy \and Harvard-Smithsonian Center for Astrophysics, Mail
  Stop 72, 60 Garden Street, Cambridge, MA 02138 \and University of
  Vienna, T\"urkenschanzstrasse 17, 1180 Vienna, Austria}
\date{Received ***date***; Accepted ***date***}

\abstract{%
  In this paper we present a novel method to identify and characterize
  stellar clusters deeply embedded in a dark molecular cloud.  The
  method is based on measuring stellar surface density in wide-field
  infrared images using star counting techniques. It takes advantage
  of the differing $H$-band luminosity functions (HLFs) of field stars
  and young stellar populations and is able to statistically associate
  each star in an image as a member of either the background stellar
  population or a young stellar population projected on or near the
  cloud.  Moreover, the technique corrects for the effects of
  differential extinction toward each individual star. We have tested
  this method against simulations as well as observations. In
  particular, we have applied the method to 2MASS point sources
  observed in the Orion~A and B complexes, and the results obtained
  compare very well with those obtained from deep \textit{Spitzer} and
  \textit{Chandra} observations where presence of infrared excess or
  X-ray emission directly determines membership status for every star.
  Additionally, our method also identifies unobscured clusters and a
  low resolution version of the Orion stellar surface density map
  shows clearly the relatively unobscured and diffuse OB 1a and 1b
  sub-groups and provides useful insights on their spatial
  distribution.}  
\maketitle

%

\defcitealias{2001A&A...377.1023L}{Paper~0}
\defcitealias{2006A&A...454..781L}{Paper~I}
\defcitealias{2008A&A...489..143}{Paper~II}
\defcitealias{2010A&A...512A..67L}{Paper~III}

\section{Introduction}
\label{sec:introduction}

Embedded star clusters offer one of the best opportunities to
understand star formation.  Within these astrophysical laboratories
hundreds of stars are formed in volumes below $1 \mbox{ pc}^3$.
Overall, it is estimated that 80\%--90\% of young stellar objects
(YSOs) form in embedded clusters (\citealp{2003ARA&A..41...57L};
although the exact fraction is somewhat sensitive to the definition of
a cluster, see \citealp{2010MNRAS.409L..54B}).  It is now established
that the fate of these objects is directly linked to the evolution of
the molecular gas, which is responsible for most of the gravitational
potential that binds the stars in the cluster.  Therefore, it is
important to study these objects in their early phases, before they
are subject to infant mortality \citep{1984ApJ...285..141L,
  2001MNRAS.323..988G}.

\begin{figure}
  \centering
  \includegraphics[width=\hsize]{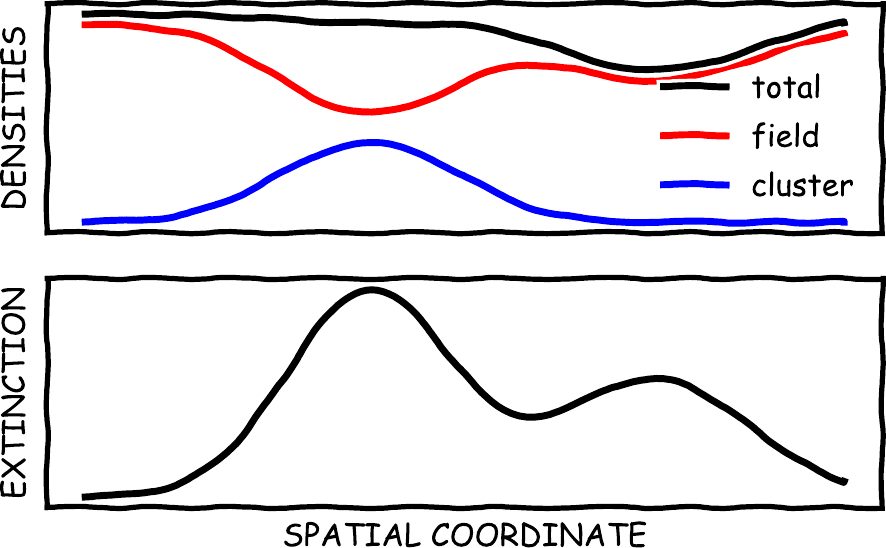}
  \caption{A one-dimensional sketch figure that illustrates the
    difficulties encountered in detecting embedded clusters.  The
    bottom black line shows the extinction profile of a molecular
    cloud.  As a result of the extinction, background stars in the
    field show an underdensity.  Even if an embedded cluster is
    present, the total observed star surface density does not reveal
    it, making it virtually impossible to detect the cluster without
    the use of the technique described in this paper.}
  \label{fig:1}
\end{figure}

Operationally, clusters are often defined and identified as groups of
stars whose stellar surface density exceeds that of field stars of the
same physical type (see, e.g., \citealp{1985prpl.conf..297W,
  1991ApJ...371..171L, 2000AJ....120.3139C}).\footnote{In this paper
  we focus on the optical identification of a cluster, and we
  explicitly ignore issues such as the gravitational stability of
  overdensities of stars.  Therefore, we will call ``cluster'' any
  overdensity, irrespective of its boundness.}\@ In this respect,
embedded clusters pose special problems because they are buried in
dust and gas, and therefore are often not even visible in optical
images; moreover, their shape is often elongated and clumpy,
reflecting the initial structure of the dense molecular gas
\citep{2005ApJ...632..397G, 2016AJ....151....5M}.  However, even with
infrared observations, discovering deeply embedded clusters and
measuring their basic parameters (such as surface density and size)
can still be a challenge since the associated dust extinguishes both
the cluster members and the field stars behind the cloud (see, e.g.,
\citealp{2008ApJ...672..861R}). In fact, typical optical or
near-infrared observations stellar fields around molecular clouds show
underdensities at the location of the clouds because of the effects of
extinction on the density of background stars. In such a situation the
observed cluster stellar surface density can be comparable or even
less than the unobscured field stellar surface density (see
Fig.~\ref{fig:1}).

Different authors have used different techniques to take into account
the effects of extinction in the identification and characterization
of star clusters. \citet{2000AJ....120.3139C} built density maps in
the direction of nearby molecular clouds using the 2MASS Point Source
Catalog, and corrected the effect of dust extinction using publicly
available CO maps (converted into infrared extinction with a constant
X-factor). In his technique, the extinction correction is only applied
to field stars: the cluster stellar density is obtained by 
subtracting an extinction-corrected model of the field stellar density
to the observed stellar density. Moreover, the use of CO maps and the
largely uncertain X-factor represent a primary source of error.

An opposite approach has been adopted by \citet{2013A&A...557A..29C},
who studied star clusters embedded in the Rosette nebula using 2MASS
data.  In this case, the local extinction was determined directly
from the stellar near-infrared colors, and a correction for the
effects of extinction was applied to each star. In practice, as noted
by the authors themselves, the correction applied would only be
correct for background stars, while it is used for all stars
(including the foreground, which should not be corrected, and embedded
ones, which should be partially corrected).  Things are further
complicated because molecular clouds are know to have steep gradients
in their column densities, and these, if undetected (because of
resolution effects, which is surely the case of the Rosette cloud at
the 2MASS resolution), would introduce further biases in the
extinction map and in the correction to the cluster richness.

\citet{2006A&A...445..999C} use yet another technique (similar to
\citealp{2002AJ....123.2559C}) to study the properties of IC~348 in
Perseus. The basic idea is that two independent measurements of
extinction in molecular clouds are available, one from the star color
excess (here applied to $H - K$ color), and one from the star number
counts. In absence of contaminants, both measurements should
agree. The presence of a cluster, however, will affect both the color
excess measurement (in a way that depends on the intrinsic color of
the cluster members) and the star count method (in a way that depends
on the location of the cluster within the cloud). The difference of
the two extinction estimates is a proxy for the cluster density. The
various assumptions made when applying this technique (in particular,
the degree of embedding of the cluster within the cloud) needs to be
resolved by calibrating it using independent measurements of cluster
richness: this clearly limits its application.

In this paper we present a new methodology to identify and
characterize clusters in wide-field or all sky, multi-band, infrared
imaging surveys.  The method is based on the production of
extinction-corrected maps of stellar surface density, and takes
advantage of the different $H$-band luminosity functions of embedded
clusters with respect to that of the background population.
Additionally, in contrast to the methods described above, it is able
to correct for cluster members unidentified because of extinction in a
way that takes into account the position of the cluster within the
cloud along the line of sight.  The technique is based on a rigorous
mathematical framework; this provides a clear advantage, in that we
can perform a detailed statistical analysis of the method. But the
detailed mathematical derivation of this method might not be of as
much interest to astronomers as its implementation.  However, we
stress that those readers not interested in the detailed mathematical
aspects of the derivation can still benefit from the method proposed
here, because its application is simple and straightforward:
Eq\eqref{eq:30}, with optionally a second expression for the noise
estimate, Eq.~\eqref{eq:31}.  We also provide a pseudo-code for it in
the appendix.

This paper is organized as follows. In Sect.~2 we present the general
framework and discuss the standard technique generally employed to
identify clusters.  In Sect.~3 we present our new method and we
discuss its practical implementation in Sect.~4.  Simple numerical
tests used to validate the method are described in Sect.~5, while the
results of an application to the Orion molecular complex are discussed
in Sect.~6.  Finally, we summarize the results obtained in this paper
in Sect.~7.

\section{The standard technique}
\label{sec:technique}


Traditionally, star clusters are identified as overdensities in the
distribution of stars.  Although many different options are available
to estimate the local angular density of stars $\sigma$
(see, e.g., \citealp{1963ZA.....57...47V, 1985ApJ...298...80C,
  2009ApJS..184...18G}), we consider in this paper the simple ``moving
average'' estimate\footnote{Throughout this paper we will use ``hats''
  to denote measured quantities.  Hence, $\sigma(\vec x)$ is the
  \textit{true} star density at the angular position $\vec x$, while
  $\hat\sigma(\vec x)$ is the \textit{measured} star density at the
  same position.}
\begin{equation}
  \label{eq:1}
  \hat \sigma(\vec x) = \sum_{n=1}^N W\bigl(\vec x - \vec x_n
  \bigr) \; .
\end{equation}
Here $n$ runs on the individual stars, $\{ \vec x_n \}$ are the
locations of the various stars, and $W$ is a window function,
normalized to unity:
\begin{equation}
  \label{eq:2}
  \int W(\vec x') \, \diff^2 x' = 1 \; .
\end{equation}
So, by construction, $W$ has a unit of one over area.  As a simple
example, suppose that $W$ is taken to be a top-hat function:
\begin{equation}
  \label{eq:3}
  W(\vec x') =
  \begin{cases}
    1 / (\pi r^2) & \text{if $|\vec x'| < r \; ,$} \\
    0                 & \text{otherwise} \; .
  \end{cases}
\end{equation}
In this case, the sum of Eq.~\eqref{eq:1} runs just over the $N_r$ stars
within an angular distance $r$ from the point of the estimate, and the
observed estimates reduces to
\begin{equation}
  \label{eq:4}
  \hat\sigma = \frac{N_r}{\pi r^2} \; .
\end{equation}
For a constant density of stars, this quantity is unbiased, in the
sense that the mean (ensemble average) $E$ of $\hat\sigma$ is
just the true density of stars:
\begin{equation}
  \label{eq:5}
  \E [ \hat\sigma ] \equiv \langle \hat \sigma \rangle =
  \frac{\E[N_r]}{\pi r^2} = \sigma \; ,
\end{equation}
and the associated variance is
\begin{equation}
  \label{eq:6}
  \Var[\hat\sigma] = \frac{\E[N_r]}{(\pi r^2)^2} =
  \frac{\sigma}{\pi r^2} \; .
\end{equation}
This equation shows that the error associated with the measured star
density decreases as $N_r^{-1/2}$: therefore, if it
is known that the density of stars is constant, to determine its value
it is sensible to use relatively large window functions $W$.

More generally, if the star density is variable, the average measured
density can be shown to be a convolution of the true density with the
window function $W$:
\begin{equation}
  \label{eq:7}
  \E [ \hat\sigma ] (\vec x) = \int
  W(\vec x - \vec x') \sigma(\vec x') \, \diff^2 x' \; ,
\end{equation}
and the associated two point correlation function is
\begin{align}
  \label{eq:8}
  \Cov & [\hat\sigma] (\vec x, \vec x') \notag\\
  & {} \equiv 
  \E \Bigl[ \bigl( \hat\sigma(\vec x) - 
  \E [\hat\sigma] (\vec x) \bigr) \bigl(
  \hat\sigma(\vec x') -  
  \E [ \hat\sigma ] (\vec x') \bigr) \Bigr]
  \notag\\
  & {} = \int W(\vec x - \vec x'') W(\vec x' - \vec x'') 
  \sigma(\vec x'') \, \diff^2 x'' \; .
\end{align}
Equation~\eqref{eq:7} shows that $W$ sets the scale for the resolution
of the density map; similarly, Eq.~\eqref{eq:8} shows that points
close enough in the density map will have correlated noise.
Therefore, if one aims at finding changes in the star density (such as
a star cluster), the window function should have a scale that is
comparable or smaller than the typical size of the variations of the
star density.  However, this is in tension with the noise properties
of Eq.~\eqref{eq:8}, because a small window function implies a large
noise.

In order to be able to detect a star cluster, the measured density of
stars at the location of the cluster must differ from the average
density much more than the standard deviation of $\hat\sigma$.  Hence,
the quantity $\Cov[\hat\sigma](\vec x, \vec x)$ sets the sensitivity
of the cluster finding algorithm.  In general the true density $\sigma
= \sigma_\mathrm{field} + \sigma_\mathrm{cluster}$ is the sum of the
field density and of the cluster density, and in some cases
$\sigma_\mathrm{cluster} \ll \sigma_\mathrm{field}$.  In these
conditions the error is dominated by the shot-noise due to the field
star population.  In reality, many other effects can prevent the
discovery and characterization of star clusters:
\begin{itemize}
\item Extinction by dark nebul\ae, which reduces the surface density
  of background sources;
\item The galactic structure, which induces smooth large-scale
  variations;
\item Differences in the sensitivity across a mosaic image due to
  changes in the observational conditions;
\item Other systematical observational effects, such halos produced by
  bright objects within the image;
\end{itemize}

\subsection{Extinction correction}
\label{sec:extinct-corr}

Among the effects listed above, the first one is particularly
important for young clusters, since these objects tend to be deeply
embedded and thus can escape a detection; additionally, detected
clusters are plagued by large uncertainties in their physical
properties (number of stars, mass, and size). For this reason,
\citet{2013A&A...557A..29C} developed a technique to perform a simple
correction of the results produced by Eq.~\eqref{eq:1}. They noted
that the density of background stars observed through a molecular
cloud decreases by a factor $10^{-\alpha A}$, where $\alpha$ is the
exponential slope of the number counts and $A$ is the extinction, both
in the band used for the observation.  Therefore, to account for the
undetected stars one can just multiply the local density estimate
$\hat \sigma(\vec x)$ by $10^{\alpha A(\vec x)}$, where $A(\vec x)$ is
the local estimate extinction (i.e., the extinction derived from an
extinction map at the location of the $\vec x$).

The problem of this approach is that it uses the same correction
factor for foreground stars, embedded objects, and background stars,
which generally results in an overestimate of the local corrected star
density. Additionally, the same correction is applied to young stellar
objects (YSOs) and field stars, which however have number count
distributions very different. This leaves a large uncertainty on the
corrected $\hat \sigma(\vec x)$ and, ultimately, on the
characterization of each cluster.

\section{Maximum-likelihood approach}
\label{sec:maxim-likel-appr}

\subsection{Constant extinction, no weighting}
\label{sec:constant-extinction}

As discussed in the previous section, the error associated to
$\hat\sigma$ is often dominated by shot noise due to the field star
population.  However, as mentioned, YSOs have photometric properties
that differ significantly from these of field stars: the former have a
$H$-band luminosity function (hereafter HLF) that can be approximated
by a Gaussian distribution \citep{2002ApJ...573..366M}, while the
latter have a HLF that follows, up to $H \sim \SI{18}{mag}$, an
exponential with slope $\alpha \sim 0.33$ (see, e.g.,
\citealp{2016A&A...587A.153M}).  This suggests that we might use
suitable cuts in the $H$-band to reduce the number of field stars,
while essentially retaining all YSOs, thus gaining in terms of noise.
However, this naive procedure is difficult to apply, since both YSOs
and field stars are also affected by extinction, which would change
the shape and the peaks of the respective HLFs.  Therefore, we are
forced to use a more systematic approach.

Let us first model the observed $H$-band luminosity function for a set
of angularly close stars (i.e., in a patch of the sky).  We suppose
that the true, unextinguished HLF can be described as a mixture of $L$
different HLFs, corresponding to different stellar populations (in the
situation considered later on in this paper we will use just two
components, one for the field stars and one for the YSOs, but
deeper observations might require the inclusion of a third component
corresponding to galaxies and following exponential number counts with
a slope of $0.6$; see \citealp{1926ApJ....64..321H}). The observed HLF
differs from the true one because of extinction and because of
incompleteness in the faint end; additionally, we will also have
photometric errors, but generally these will be small compared to the
typical width of the various HLF components, and for the sake of
simplicity are therefore neglected here.\footnote{Photometric errors
  can be easily included in Eq.~\eqref{eq:9} by replacing $p_i(m - A)$
  there with the convolution of this function with the photometric
  uncertainty.  In presence of different uncertainties for each star,
  one will need to specialize Eq.~\eqref{eq:9} to each star.  A
  similar technique can be used to include the effects of errors on
  the extinction measurements.}\@ We can thus write finally the
observed HLF, measured in units of stars per area per magnitude bin, as
\begin{equation}
  \label{eq:9}
  \sigma(m) = c(m) \sum_{i=1}^L \sigma_i p_i(m - A) \; ,
\end{equation}
where $A$ is the $H$-band extinction,\footnote{In this section we will
  take the extinction to be identical for a set of angularly close
  stars; we will then relax this assumption and use the individual
  extinction in the direction of each star.} $c(m)$ is the
completeness function (i.e.\ the probability to detect a star with
apparent magnitude $m$), $p_i(m)$ is the probability distribution for
the $i$-th component of the HLF, and $\sigma_i$ is a coefficient that
indicates the predominance of the $i$-th component in the HLF.  In
order to give $\sigma_i$ a simple interpretation, it is useful to take
$p_i$ suitably normalized.  In particular, we assume that
\begin{equation}
  \label{eq:10}
  \int c(m) p_i(m) \, \diff m = 1 \; .
\end{equation}
With this choice, $\sigma_i$ can be identified as the observed
angular density of stars for the $i$-th component in absence of
extinction.

Our goal is to infer the parameters $\{ \sigma_i \}$ from a set of
observed $H$-band magnitudes $\{ m_n \}$ in a patch of the sky of area
$S$; we will then repeat this operation in different directions in
order to build maps of densities for the various components.  This
will allow us to identify overdensities in each component, such as in
the YSO one.  To this purpose, we use Eq.~\eqref{eq:9} to write
the log-likelihood following the prescriptions of
\citet{2013A&A...559A..90L}.  The entire analysis is done in the
magnitude space, by using $S \sigma$ as the probability distribution
density for the magnitudes:
\begin{align}
  \label{eq:11}
  \ln \mathcal{L} = {} & \sum_{n=1}^N \ln S \sigma(m_n) - \int
  S \sigma(m) \, \diff m \notag\\
  {} = {} & \sum_{n=1}^N \biggl[ \ln S c(m_n) \sum_{i=1}^L
  \sigma_i p_i(m_n - A) \biggr] \notag\\
  & {} - \int S c(m) \sum_{i=1}^L \sigma_i p_i(m - A) \, \diff m
  \; .
\end{align}
This likelihood can be used in Bayes' theorem to infer a posterior
probability for the densities $\{ \sigma_i \}$.  More simply, we can
just find the most likely parameters $\{ \sigma_i \}$ using a maximum
likelihood approach.  To this purpose, we consider
\begin{equation}
  \label{eq:12}
  \frac{\partial \ln \mathcal{L}}{\partial \sigma_i} = \sum_{n=1}^N
  \frac{c(m_n) p_i(m_n - A)}{\sigma(m_n)} - S \! \int c(m) p_i(m-A) \,
  \diff m \; .
\end{equation}
The maximum likelihood solution is given by the set of densities
$\sigma_i$ that maximize $\mathcal{L}$ or, equivalently, $\ln
\mathcal{L}$, i.e.\ by the zero of Eq.~\eqref{eq:12}; that is, we need
to solve
\begin{equation}
  \label{eq:13}
  \sum_{n=1}^N \frac{c(m_n) p_i(m_n - A)}{\sigma(m_n)} = S \! \int
  c(m) p_i(m-A) \, \diff m \; .
\end{equation}
Unfortunately, the solution of this equation for $L > 1$ cannot be
provided in analytic form, and must be obtained by numerical methods.
We will discuss below in Sect.~\ref{sec:implementation} simple ways to
find it.

We can obtain an estimate of the errors associated to the measurements
$\sigma_i$ from the Fisher information matrix (see, e.g.,
\citealp{ModStat}), that we recall is defined as
\begin{equation}
  \label{eq:14}
    I_{ij} = \E \left[ \frac{\partial \ln \mathcal{L}}{\partial
      \sigma_i} \frac{\partial \ln \mathcal{L}}{\partial
      \sigma_j} \right] = - \E \left[ \frac{\partial^2 \ln
      \mathcal{L}}{\partial \sigma_i \, \partial \sigma_j} \right] \; .
\end{equation}
The Fisher information matrix is related to the minimum
covariance matrix that can be attained by an unbiased estimator, as
provided by the Cram\'er-Rao bound:
\begin{equation}
  \label{eq:15}
  \Cov[\sigma] \ge I^{-1} \; .
\end{equation}
Since the maximum-likelihood estimator is asymptotically efficient
(i.e.\ it attains the Cram\'er-Rao bound when the sample size tends to
infinity) and the resulting errors on $\hat\sigma$ tend to a
multivariate Gaussian distribution, it is interesting to obtain an
analytic result for the information matrix.

The Fisher information matrix can be readily evaluated from our data
using Eq.~(12) of \citet{2013A&A...559A..90L}:
\begin{align}
  \label{eq:16}
  I_{ij} = S \! \int \frac{c^2(m) p_i(m-A) p_j(m-A)}{\sigma(m)} \,
  \diff m \; .
\end{align}
This relation is interesting from several points of view.  First, note
that the Fisher matrix contains elements outside the diagonal, unless
the probability distributions for the various components do not
overlap, i.e.\ have non-intersecting support: this would basically
mean that we could immediately tell to which component belongs a star
from its unextinguished magnitude.  Second, note that all elements of
the matrix are non-negative and therefore, in case of $L=2$
components, $\Cov[\sigma] \simeq I^{-1}$ will have non-positive
elements off-diagonal: in other words, the measurements of the two
densities $\sigma_1$ and $\sigma_2$ will be affected by a negative
correlation.  This is expected and corresponds to a classification
error: if $p_1$ overlaps with $p_2$, we cannot uniquely associate
stars to each different population and in general an overestimate of
one population is done at the expenses of an underestimate of the
other population.

It is instructive to consider the form of the information matrix in
the special case where $L=1$.  When just one component is
used, then $I$ is a scalar and it reduces to
\begin{equation}
  \label{eq:17}
  I = \frac{S}{\sigma} \int c(m) p(m-A) \, \diff m =
  \frac{S^2}{\E[N]} \; ,
\end{equation}
where $\E[N] = S \sigma$ is the average number of stars observed in
the area $S$.  Its inverse, $I^{-1}$, is therefore $\E[N] /
\sigma^2$, as expected from a simple Poisson statistics.

With $L \ge 2$, in principle we can encounter cases where the Fisher
information matrix is singular.  Given the analytic form of $I$, this
happens in particular when the two components have the exact same
probability distributions within the support of $c(m)$, i.e.\ $c(m)
p_i(m) = c(m) p_j(m)$: in this case, the corresponding rows (or
columns) of $I$ are identical.  In such a situation, clearly, it
is virtually impossible to classify objects as one of the two
degenerate components, and therefore the uncertainty on the respective
densities $\sigma_i$ and $\sigma_j$ are infinite.

For completeness, we also report the expected maximum value of the
log-likelihood
\begin{equation}
  \label{eq:18}
  \E[\mathcal{L}] = \frac{L}{2} + S \! \int \sigma(m) \bigl[ \ln
  S \sigma(m) - 1 \bigr] \, \diff m \; ,
\end{equation}
and the associated variance
\begin{equation}
  \label{eq:19}
  \Var[\mathcal{L}] = S \! \int \sigma(m) \ln^2 S \sigma(m) \,
  \diff m \; .
\end{equation}
These equations can be used to verify that the chosen model
\eqref{eq:9} can account well for the data.

\subsection{Differential extinction and spatial weight}
\label{sec:diff-extinct-spat}

So far, we have assumed that all stars are subject to the same
extinction $A$; moreover, we have not weighted stars depending on
their angular position as done in Eq.~\eqref{eq:1}.  In this section
we intend to remove these two limitations and consider the full
problem. 

The simpler approach to include the effects of differential extinction
is to consider the \textit{joint} density in magnitudes and positions.
In this framework, we can rewrite Eq.~\eqref{eq:11} for log-likelihood
as
\begin{align}
  \label{eq:20}
  \ln \mathcal{L} = {} & \sum_{n=0}^N \ln \sigma(m_n, \vec x_n)
  - \int \diff m \int \diff^2 x' \, \sigma(m, \vec x') \; .
\end{align}
In this equation the quantity $\sigma(m, \vec x')$ represents the
predicted density of stars with magnitude $m$ at the location $x'$.
Similarly to Eq.~\eqref{eq:9}, we write this quantity as a mixture of 
different densities, corresponding to different stellar populations:
\begin{equation}
  \label{eq:21}
  \sigma(m, \vec x') = c(m) \sum_{i=1}^L \sigma_i(\vec x')
  p_i \bigl( m - A(\vec x') \bigr) \; ,
\end{equation}
where $\sigma_i(\vec x')$ represents the density of stars of class $i$
at the sky position $\vec x'$.  In order to proceed, we need to model
these densities.  A simple and consistent way of doing this, is to
suppose that $\ln \sigma_i(\vec x')$ can be written as the weighted
sum of two terms: one, $\ln \sigma(\vec x)$, associated to the density
at the point of interest $\vec x$ (the point where we intend to
evaluate the local densities of stellar populations); and one, $\ln
\tau_i(\vec x')$, which parametrizes the local changes of the densities.
As a result, we write
\begin{equation}
  \label{eq:22}
  \sigma_i(\vec x') = \bigl( \sigma_i(\vec x)
  \bigr)^{\omega(\vec x - \vec x')} \bigl( \tau_i(\vec x')
  \bigr)^{1 - \omega(\vec x - \vec x')} \; .
\end{equation}
The function $\omega$ describes the spatial correlation between
densities at different positions and plays a central role in
identifying which stars contribute to the density estimate at $\vec
x$.

We can now insert Eqs.~\eqref{eq:22} and \eqref{eq:21} in
Eq.~\eqref{eq:20} and find the maximum likelihood solution over
$\sigma_i(\vec x)$.  Calling $A_n \equiv A(\vec x_n)$ the extinction
suffered by the $n$-th star, we find
\begin{align}
  \label{eq:23}
  & \frac{\partial \ln \mathcal{L}}{\partial \sigma_i(\vec x)} =
  \sum_{n=1}^N \omega(\vec x - \vec x_n) \frac{p_i (m_n - A_n)
    \sigma_i(\vec x_n) / \sigma_i(\vec x)}%
  {\sum_j \sigma_j(\vec x_n) p_j(m_n - A_n)}
  \notag\\
  & \quad - \int \diff m \, c(m) \int \diff^2 x' \, \omega(\vec x')
  p_i\bigl( m - A(\vec x') \bigr) \sigma_i(\vec x') / \sigma_i(\vec x)
  \; .
\end{align}
We now make an important assumption.  Because of the form of the
parametrization \eqref{eq:22}, a solution for the maximum likelihood
can only be found if we specify the functional form of the functions
$\tau_i(\vec x')$.  However, these functions are truly unknown, since
they parametrize the local changes of the various densities, which in
turn depend on the local density map.  Using a statistical approach,
however, it is natural to assume that these functions have the same
distribution of $\sigma_i(\vec x)$, the quantity we are interested
in.  As a result, if we take an average of Eq.~\eqref{eq:23}, terms
such as $\sigma_i(\vec x') / \sigma_i(\vec x)$ cancel out:
\begin{align}
  \label{eq:24}
  & \frac{\partial \ln \mathcal{L}}{\partial \sigma_i(\vec x)} =
  \sum_{n=1}^N \omega(\vec x - \vec x_n) \frac{p_i (m_n - A_n)}{\sum_j
    \sigma_j(\vec x) p_j(m_n - A_n)} \notag\\
  & \quad - \int \diff m \, c(m) \int \diff^2 x' \, \omega(\vec x')
  p_i\bigl( m - A(\vec x') \bigr) \; .
\end{align}

Before proceeding, it is useful to consider the solution of the
maximum likelihood approach in the simple case where there is a single
population of stars and where the extinction vanishes, $A(\vec x') =
0$.  We find in this case
\begin{equation}
  \label{eq:25}
  \frac{\partial \ln \mathcal{L}}{\partial \sigma(\vec x)} =
  \sum_{n=1}^N \frac{\omega(\vec x - \vec x_n)}{\sigma(\vec x)} -
  \int \diff^2 x' \, \omega(\vec x') \; ,
\end{equation}
where we have used the normalization property \eqref{eq:10}.  The
solution of this equation is immediately found as
\begin{equation}
  \label{eq:26}
  \sigma(\vec x) = \sum_{n=1}^N W(\vec x - \vec x_n) \; ,
\end{equation}
where
\begin{equation}
  \label{eq:27}
  W(\vec x) = \frac{\omega(\vec x)}{\int \omega(\vec x') \, \diff^2
    x'}
\end{equation}
We have therefore recovered Eq.~\eqref{eq:1}, with the correct
normalization \eqref{eq:2} for the weight $W$.

In the general case, the maximum-likelihood solution of
Eq.~\eqref{eq:24} must be obtained numerically.  The errors associated
to the solutions can be estimated using the Fisher matrix.  In our
case, we can obtain the Fisher matrix from
\begin{align}
  \label{eq:28}
  I_{ij}(\vec x) = {} & \iint \diff m \, \diff^2 x' \, \frac{1}{\sigma(m,
    \vec x')} \frac{\partial \sigma(m, \vec x')}{\partial
    \sigma_i(\vec x)} \frac{\partial \sigma(m, \vec x')}{\partial
    \sigma_j(\vec x)} \notag \\ 
  {} = {} & \iint \diff m \, \diff^2 x' \, c^2(m) \omega^2(\vec x -
  \vec x') \times \notag\\
  & \qquad \frac{p_i \bigl( m - A(\vec x') \bigr) p_j \bigl( m -
    A(\vec x') \bigr)}{\sigma(m,
    \vec x')} \frac{\sigma_i(\vec x') \sigma_j(\vec x')}{\sigma_i(\vec
    x) \sigma_j(\vec x)} \; .
\end{align}
As before, we can replace $\vec x$ to $\vec x'$ in all arguments of
$\sigma_i$, with the justification that the statistical properties of
$\sigma$ are invariant upon translation.  We will discuss in the next
section useful expressions to evaluate this quantity in practical
cases.

\begin{table*}[bt]
  \centering
  \begin{tabular}{lcccc}
    \toprule
    Function & Support & $\omega(\vec x)$ & $W(\vec x)$ & 
    Eff. area\\
    \midrule
    Top-hat & $\vec |\vec x| \le s$ & $1$ & $1/(\pi s^2)$ & $\pi s^2$ \\
    Conic & $\vec |\vec x| \le s$ & $2 (1 - |\vec x|/s)$ & $3 (1 - |\vec
    x|/s) / (\pi s^2)$ & $2 \pi s^2 / 3$ \\
    Parabolic & $\vec |\vec x| \le s$ & $3 (1 - |\vec x|^2/s^2) / 2$ &
    $2 (1 - |\vec x|^2/s^2) / (\pi s^2)$ & $3 \pi s^2 / 4$ \\
    Gaussian & $\R^2$ & 2 $\exp\bigl( - |\vec x|^2 / 2 s^2 \bigr)$ &
    $\exp\bigl( - |\vec x|^2 / 2 s^2 \bigr) / (2 \pi s^2)$ & 
    $4 \pi s^2$ \\
    \bottomrule
  \end{tabular}
  \caption{Different two-dimensional spatial functions $\omega(\vec
    x)$ and corresponding weight functions $W(\vec x)$.  All functions
    considered here are axisymmetric and include a spatial scale $s$.}
  \label{tab:1}
\end{table*}

\subsection{Implementation}
\label{sec:implementation}

The algorithm proposed in this paper is essentially the search of the
solutions of Eq.~\eqref{eq:24} as a function of the star densities $\{
\sigma_i \}$ corresponding to the various components or star
populations.  The same procedure must be applied to different patches
of the sky, so that maps of the star densities can be obtained.
These, in turn, will allow us to identify and characterize star
clusters, and in particular embedded ones.

Although the practical implementation of the algorithm follows this
schema, a number of technical and theoretical aspects must be
correctly addressed in order to optimize the detection and make the
technique efficient.

First, we note that a simple way to obtain the (positive) solutions of
Eq.~\eqref{eq:24} is through the use of a recursive formula.  To this
purpose, multiply both members of this equation by $\sigma_i(\vec x)$,
and solve for this quantity, thus obtaining the expression
\begin{equation}
  \label{eq:29}
  \sigma_i(\vec x) \leftarrow \frac{\displaystyle
    \sum_{n=1}^N \omega(\vec x - \vec x_n) \frac{\sigma_i(\vec x)
      p_i(m_n - A_n)}{\sum_{j=1}^L \sigma_j(\vec x) 
      p_j(m_n - A_n)}}{\displaystyle
    \int \diff m \, c(m) \int \diff^2 x' \, \omega(\vec x')
    p_i\bigl( m - A(\vec x') \bigr)} \; .
\end{equation}
Unfortunately, we are unable to use this equation because we only know
the extinction at the locations of the stars $A_n \equiv A(\vec x_n)$:
this prevents us from evaluating the integral over $\diff x^2$ in the
denominator.  We can, however, move the denominator inside the sum,
and evaluate the second integral by replacing $A(\vec x)$ with $A(\vec
x_n)$, the extinction at the direction of each star.  Additionally,
using the same argument that has been employed in Eq.~\eqref{eq:23},
that is the similarity between $\sigma$ and $\tau$, it is convenient
to replace $\sigma_i(\vec x)$ with $\sigma_i(\vec x_n)$ in the
numerator.  This procedure leads to the iteration
\begin{equation}
  \label{eq:30}
  \sigma_i(\vec x) \leftarrow \sum_{n=1}^N \frac{\displaystyle
    W(\vec x - \vec x_n) \frac{\sigma_i(\vec x_n)
      p_i(m_n - A_n)}{\sum_{j=1}^L \sigma_j(\vec x_n)
      p_j(m_n - A_n)}}{\int \diff m \, c(m) p_i(m - A_n)} \; .
\end{equation}
Equation~\eqref{eq:30} is the solution proposed in this paper to
estimate the local density of stars.  As indicated by the left arrow
symbol, we can obtain the set of values $\{ \sigma_i \}$ by starting
with some arbitrary (positive) values for these quantities, and then
by calculating updated values of $\sigma_i$ by applying
Eq.~\eqref{eq:30}.  The convergence is usually reached within a few
tens of iterations.

Note that Eq.~\eqref{eq:30} has a simple interpretation.  Let us
ignore for a moment the weight $W$, i.e.\ let us assume that all stars
have the same weight.  The sum in Eq.~\eqref{eq:30} is carried out
over all stars in the patch of the sky, but each star is counted only
partially (i.e., contributes with a term between zero and unity in the
sum): precisely, each star contributes by the computed probability
that the star be associated with the $i$-th component.  The way this
probability is computed is actually a simple application of Bayes'
theorem, where $p_i(m_n - A_n)$ plays the role of the likelihood,
$\sigma_i(\vec x_n)$ is proportional to the prior that the star is of
class $i$, and the sum over $j$ in the denominator is proportional to
the evidence.  The result of the sum of these terms is divided by the
by the result of the integral: this is a correcting factor that takes
into account the fact that, because of extinction and incompleteness,
we will miss a fraction of stars.  Note also that Eq.~\eqref{eq:30}
can be also considered as a special case of a $K$-means soft
clustering algorithm where the only unknown quantities are the classes
$\sigma_i$ \citep[see][]{MacKay}.

Before proceeding, it is useful to recall the hypotheses of this
algorithm and its strengths.  First, we assume that we have some
knowledge of the $H$-band luminosity function for the various
populations of stars that are likely to be present in the field. In
practice, we will generally use two probabilities, one for the field
stars, and one for the YSOs. Second, we assume that we have measured
the extinction $A_n$ of each star: note that this is \textit{not} the
average extinction at the location of the star, which might be very
different because of perspective effects: for example, a foreground
star in front of a cloud would have $A_n \simeq 0$, while the average
extinction would be significant. This way, the algorithm can directly
account for foreground contamination: foreground stars will not be
corrected in their counts, since the integral in the denominator of
Eq.~\eqref{eq:30} will evaluate to unity. Similarly, stars within
molecular clouds will be corrected only for the amount of material
that is really in front of them.

Finally, we stress that the iterative procedure proposed here only
find positive solutions for the values $\sigma_i$.  Although
reasonable, nevertheless this choice inevitably introduces biases in
the results: for example, in a region where no YSO is present,
because of errors we will still measure small positive values for the
density of YSOs.  However, numerical tests have shown that the
bias amount is limited; moreover, a reduction of the bias is
associated to a large increase in the scatter.  Therefore, we will
force the $\sigma_i$ to be positive and use Eq.~\eqref{eq:30} for the
solution.

The uncertainties associated to Eq.~\eqref{eq:30} can be computed from
the Fisher matrix.  For practical applications, it is convenient to
rewrite Eq.~\eqref{eq:28} by replacing the integrals over $\diff m$
and $\diff^2 x'$ with a sum over the observed objects.  This leads to
the approximated Fisher matrix expression
\begin{equation}
  \label{eq:31}
  I_{ij} = \sum_{n=1}^N \frac{\omega^2(\vec x - \vec x_n) c^2(m_n)
    p_i(m_n - A_n) p_j( m_n - A_n)}{\sigma^2(m_n, \vec x_n)} \; . 
\end{equation}
In this equation, we take the spatial function $\omega$ to be
normalized such that
\begin{equation}
  \label{eq:32}
  \int \omega(\vec x') \, \diff^2 x' = \int \omega^2(\vec x') \,
  \diff^2 x' \; ;
\end{equation}
that is, in terms of $W$,
\begin{equation}
  \label{eq:33}
  \omega(\vec x) = \frac{W(\vec x)}{\int W^2(\vec x') \, \diff^2 x'}
  \; .
\end{equation}
Table~\ref{tab:1} reports a few common choices for the spatial
function $\omega(\vec x)$ and the corresponding weight function
$W(\vec x)$, both correctly normalized.  As usual, the covariance
matrix associated with the measurements of the densities $\sigma_i$
can be computed from the inverse of the Fisher matrix, $I^{-1}$.

\section{Simulations}
\label{sec:simulations}

\begin{table}[tb]
  \centering
  \begin{tabular}{ccccc}
    \toprule
    $\sigma_1$ & $\sigma_2$ & $\langle (I^{-1})_{11} \rangle$ & 
    $\langle (I^{-1})_{12} \rangle$ & $\langle (I^{-1})_{22} \rangle$ \\
    $\langle \hat\sigma_1 \rangle$ & $\langle \hat\sigma_2 \rangle$ &
    $\Var[\sigma_1]$ & $\Cov[\sigma_1,\sigma_2]$ & $\Var[\sigma_2]$ \\
    \midrule
    $5.00$ & $20.00$ & $33.38$ & $-19.13$ & $35.27$ \\
    $5.30$ & $19.87$ & $21.37$ & $-12.09$ & $29.60$ \\
    \midrule
    $20.00$ & $ 5.00$ & $47.44$ & $-19.02$ & $22.34$ \\
    $19.79$ & $ 5.21$ & $43.31$ & $-13.83$ & $16.33$ \\
    \midrule
    $10.00$ & $10.00$ & $47.44$ & $-19.02$ & $22.34$ \\
    $10.05$ & $ 9.95$ & $30.63$ & $-13.97$ & $21.04$ \\
    \bottomrule
  \end{tabular}
  \caption{Summaries of the results of simulations.}
  \label{tab:2}
\end{table}

\begin{figure}[t]
  \centering
  \includegraphics[width=\hsize]{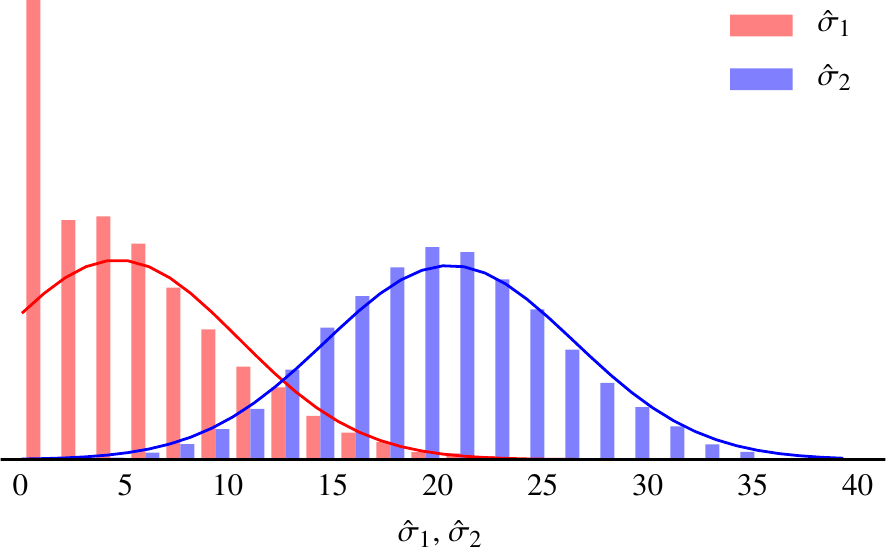}
  \caption{The distributions of measured densities in a simulation
    with $\sigma_1 = 5$ and $\sigma_2 = 20$ (histogram), together with
    the predicted Gaussian distribution obtained according to the
    Fisher matrix evaluated from Eq.~\eqref{eq:31}.  The excess of
    small values of $\hat\sigma_1$ is due to the constraint that
    $\hat\sigma \ge 0$ imposed by the algorithm.}
  \label{fig:3}
\end{figure}

\begin{figure}[t]
  \centering
  \includegraphics[width=\hsize]{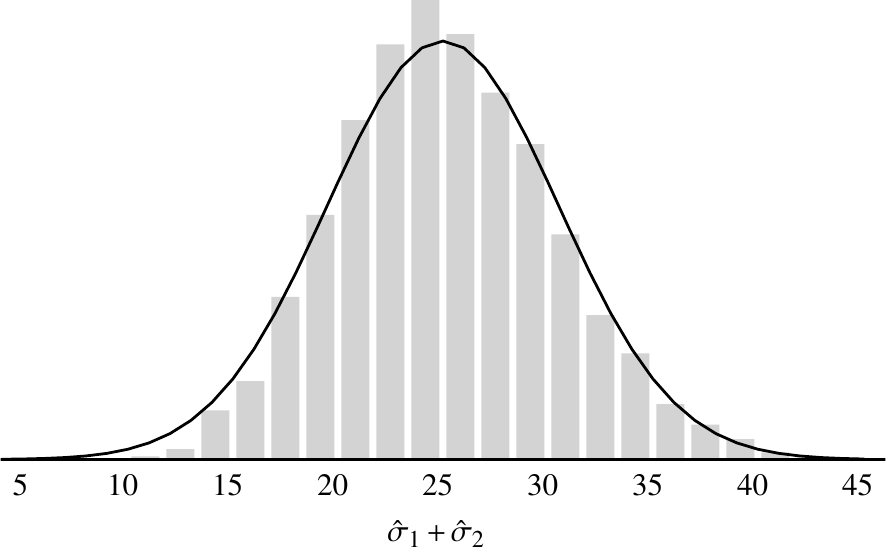}
  \caption{The distribution of measured total density $\hat\sigma_1 +
    \hat\sigma_2$ in a simulation with $\sigma_1 = 5$ and $\sigma_2 =
    20$, together with the predicted Gaussian distribution (derived
    from the Fisher matrix).  The distribution is essentially
    unbiased; moreover, because of the anticorrelation between
    $\hat\sigma_1$ and $\hat\sigma_2$, the total density has
    significantly less scatter than both $\hat\sigma_1$ and
    $\hat\sigma_2$.}
  \label{fig:4}
\end{figure}

\begin{figure}[t]
  \centering
  \includegraphics[width=\hsize]{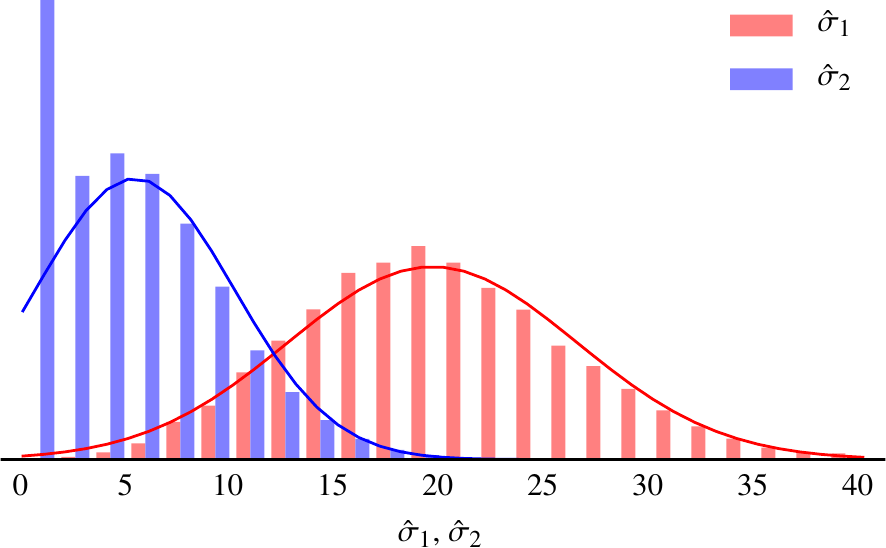}
  \caption{The distributions of measured densities in a simulation
    with $\sigma_1 = 20$ and $\sigma_2 = 5$.}
  \label{fig:5}
\end{figure}

\begin{figure}[t]
  \centering
  \includegraphics[width=\hsize]{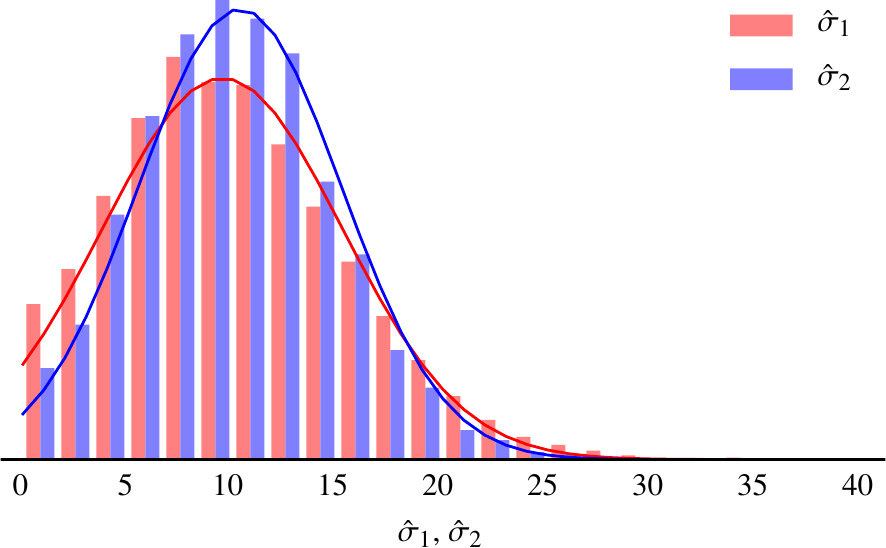}
  \caption{The distributions of measured densities in a simulation
    with $\sigma_1 = 10$ and $\sigma_2 = 10$.}
  \label{fig:6}
\end{figure}

\begin{figure}[t]
  \centering
  \includegraphics[width=\hsize]{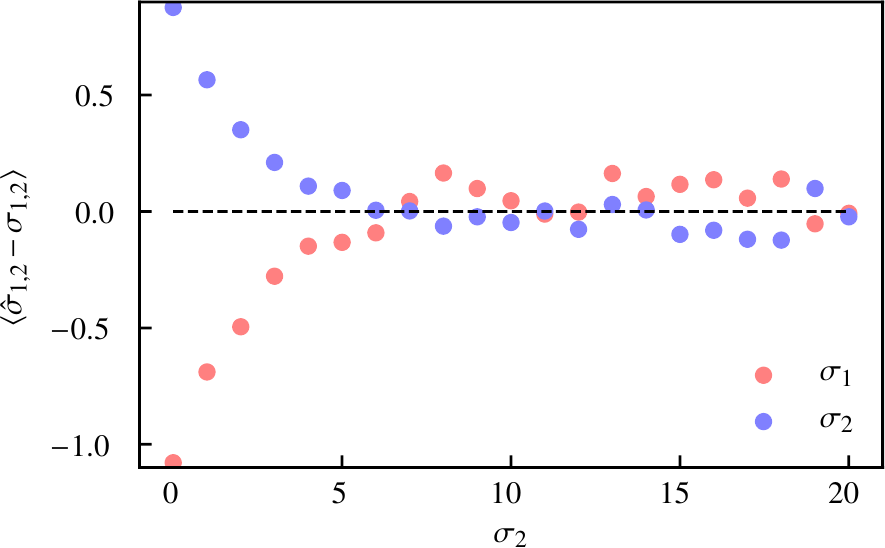}
  \caption{The biases in the measured densities for $\sigma_1 = 10$
    and $\sigma_2 \in [0, 20]$.  Note how the bias essentially
    vanishes for $\sigma_2 > 5$.}
  \label{fig:7}
\end{figure}

\begin{figure}[t]
  \centering
  \includegraphics[width=\hsize]{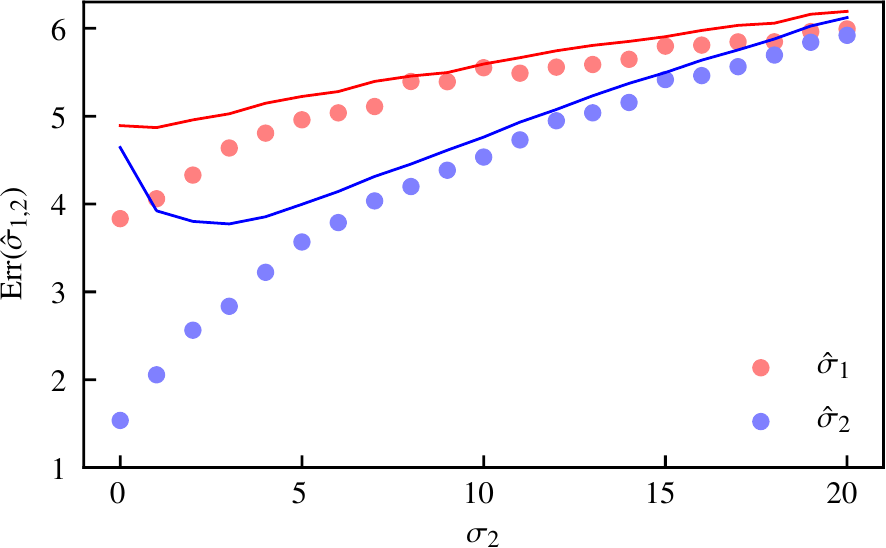}
  \caption{The scatter on $\hat\sigma_1$ and $\hat\sigma_2$ for
    $\sigma_1 = 10$ and $\sigma_2 \in [0, 20]$ (dots), together with
    the predictions obtained from Eq.~\eqref{eq:31} (lines).}
  \label{fig:8}
\end{figure}

\begin{figure}[t]
  \centering
  \includegraphics[width=\hsize]{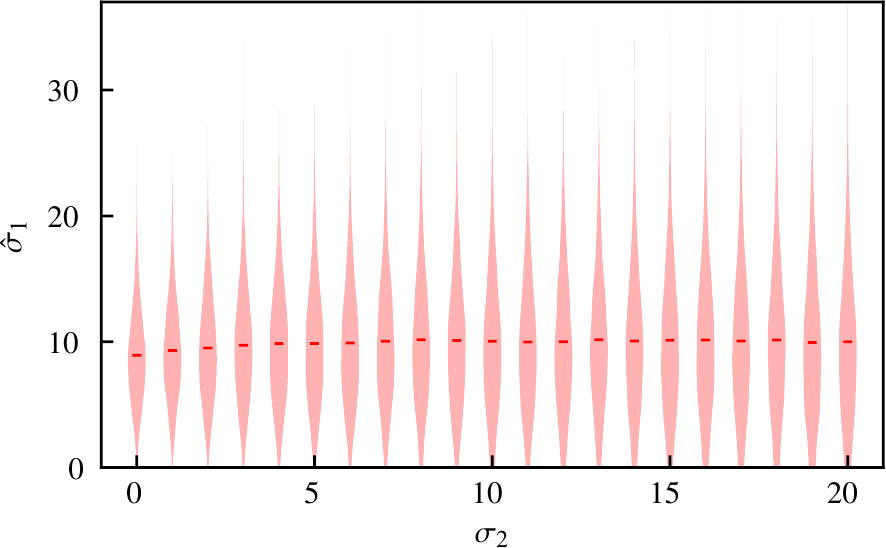}
  \caption{Violin plot showing the distribution of measured densities
    $\hat\sigma_1$ for $\sigma_1 = 10$ and $\sigma_2 \in [0, 20]$.
    Each elongated structure corresponds to a different value of
    $\sigma_2$; its width is proportional to the distribution of
    measured values of $\hat \sigma_1$, i.e.\ effectively it is a
    histogram displayed vertically. The small red dashes indicate the
    average values.}
  \label{fig:9}
\end{figure}

\begin{figure}[t]
  \centering
  \includegraphics[width=\hsize]{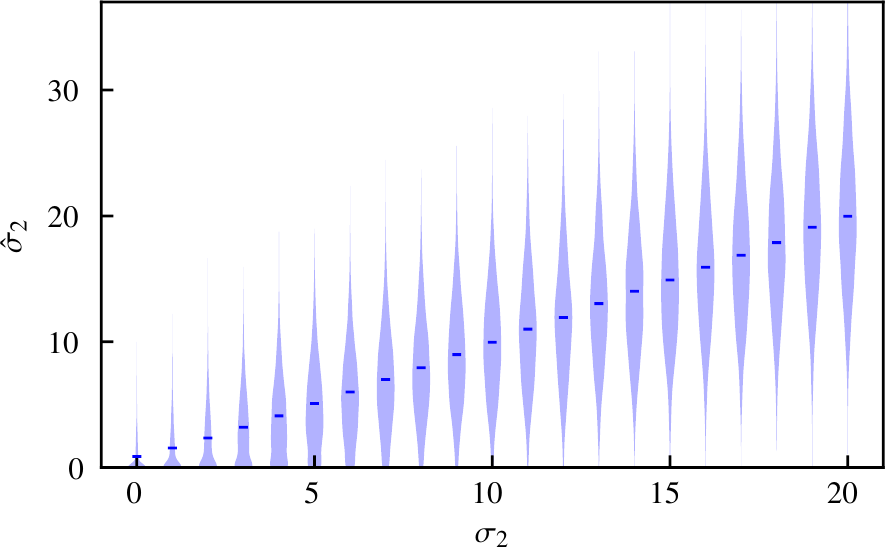}
  \caption{Violin plot showing the distribution of measured densities
    $\hat\sigma_2$ for $\sigma_1 = 10$ and $\sigma_2 \in [0, 20]$.
    The small blue dashes indicate the average values.}
  \label{fig:10}
\end{figure}
In order to test our method and verify its robustness we have
performed a set of simulations.  We have considered a small patch of
the sky with two different stellar populations: field stars, with
exponentially increasing number counts $p_1(m) \propto 10^{\alpha m}$,
with $\alpha = \SI{0.33}{mag^{-1}}$; and YSOs, with Gaussian
number counts $p_2(m) \propto \exp\bigl(-(m-m_0)^2 / 2 s^2 \bigr)$,
with $m_0 = \SI{12}{mag}$ and $s = \SI{1.65}{mag}$.  We have
distributed both populations randomly in the small patch of the sky
considered and we have added to each star a random extinction drawn
from the probability distribution $p_A(A) \propto A^{-2}$, in the
range $A \in [0.1, 2.0]\,\si{mag}$.  This choice is meant to simulate
the effects of differential extinction for objects within molecular
clouds.  Finally, we have imagined that our observations are complete
up to $\SI{15}{mag}$, and that no stars can be observed beyond that
magnitude: in other words, we have modeled the completeness function
as a Heaviside function $c(m) = H(15 - m)$.  This way, our final
catalog contains, for each star, the angular position in the sky, the
$H$-band magnitude, and the measured extinction.  Note that the
parameters used here are chosen to simulate a real situation
corresponding to the sample application of
Sect.~\ref{sec:sample-appl-orion}, that is the Orion molecular cloud
complex observed with 2MASS.

We have used these data in our algorithm, represented by
Eqs.~\eqref{eq:30} and \eqref{eq:31}.  As weight function $W(\vec x)$
we have used a Gaussian, and we have chosen the angular units so that
\begin{equation}
  \label{eq:34}
  \int W^2(\vec x') \, \diff^2 x' = \left[ \int W(\vec x') \, \diff^2
  x' \right]^2 \; .
\end{equation}
This choice guarantees that the effective area of the weight function
is unity, i.e.\ the effective number of stars used for the analysis,
in absence of extinction, would be just $\sigma_1 + \sigma_2$.   In
reality, the presence of the extinction reduces this number by a
factor that depends on the relative ratio between $\sigma_1$ and
$\sigma_2$ (typically, by $\sim 20\%$).

We have then performed different simulations, with various choices for
$\sigma_1$ and $\sigma_2$, to verify the ability of the algorithm to
recover the input densities.  Figures~\ref{fig:3}--\ref{fig:6}
show the observed distributions of $\hat\sigma_1$ and $\hat\sigma_2$,
together with the predicted ones (Gaussian distributions centered on
the true values of $\sigma_1$ and $\sigma_2$, with the variances
predicted from the Fisher matrix $I$).  In general, we can note a few
points:
\begin{itemize}
\item When one of the input densities is below $\sim 7$, there is a
  clear excess of the corresponding quantity for small measured
  values.  This is a consequence of the fact that the proposed
  algorithm only returns positive values for the densities.
\item Except for the point above, the predicted distributions
  reproduce very well the measured data.  The agreement is
  particularly good when the input densities are large.
\item Overall, the total density $\sigma_1 + \sigma_2$ is better
  constrained than the individual densities $\sigma_{1,2}$.
\end{itemize}
Figures~\ref{fig:7} and \ref{fig:8} show the biases and the errors
on the measured densities $\hat \sigma_{1,2}$ for $\sigma_1 = 10$ and
$\sigma_2$ varying in the interval $[0, 20]$.  We can see that there
is a measurable bias for $\sigma_2 < 5$, while the results are
essentially unbiased above this value.  Correspondingly, we observe in
the same range $\sigma_2 \in [0, 5]$ a measured scatter in measured
densities that is significantly smaller than what predicted from the
Fisher matrix.  For larger values of the input density the error
estimate gets closer to the measured errors, but still remains
slightly above.  This is actually good, because implies a small
overestimate of the error which will make the entire method more
robust for cluster detections (that is, it will decrease the number of
false positive).  In summary, all these simulations confirm that the
method works and that the error estimate is quite accurate.

\section{Sample application: Orion}
\label{sec:sample-appl-orion}

\begin{figure*}[t]
  \centering
  \includegraphics[width=\hsize]{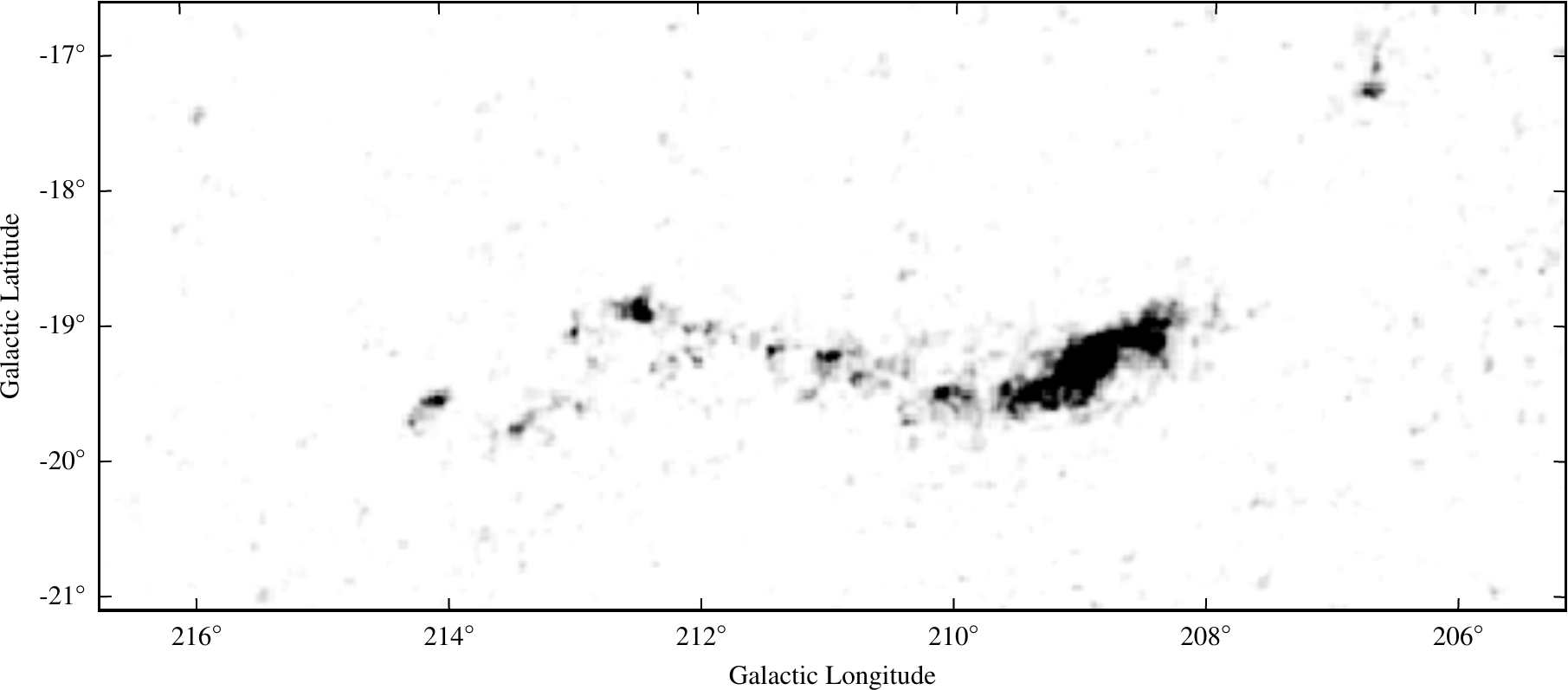}%
  \hspace{-\hsize}%
  \begin{ocg}{fig:30b2off}{fig:30b2off}{0}%
  \end{ocg}%
  \begin{ocg}{fig:30b2on}{fig:30b2on}{1}%
    \includegraphics[width=\hsize]{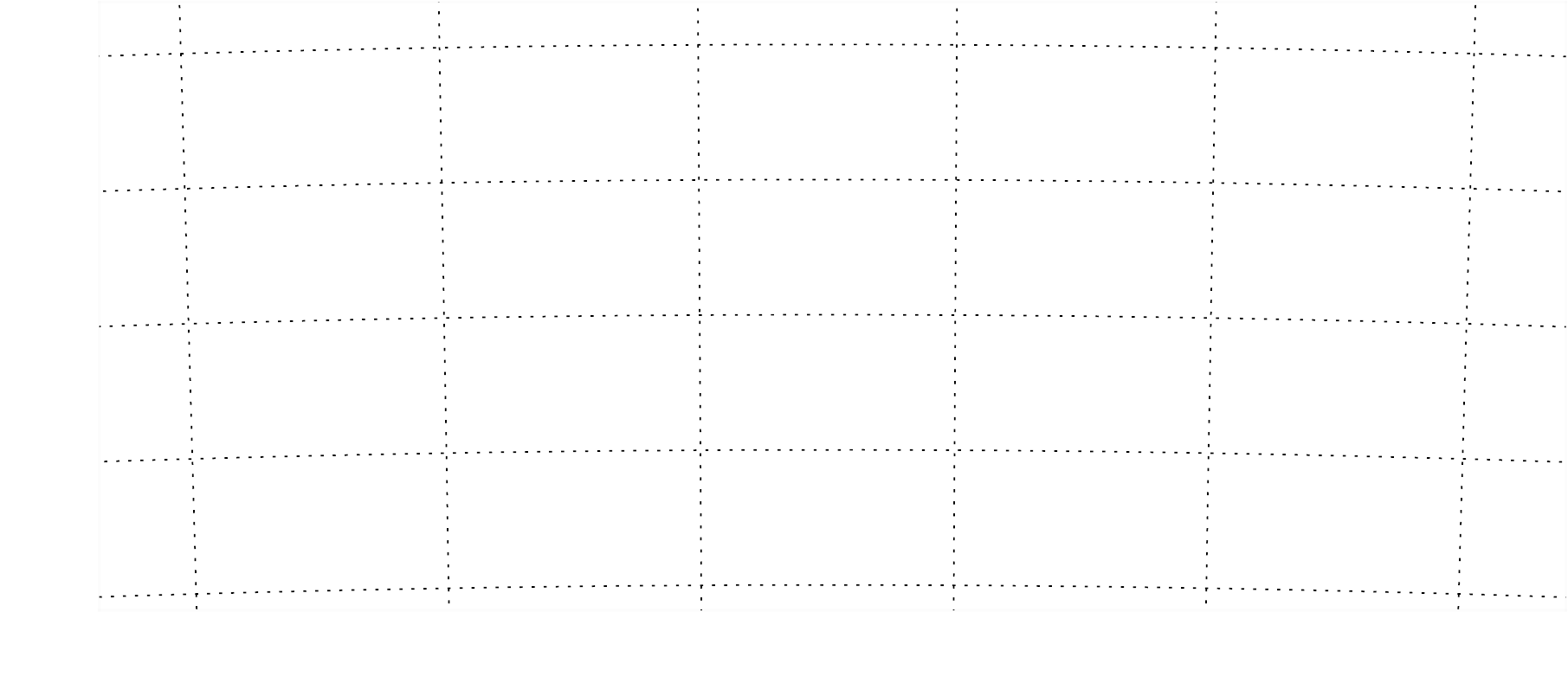}%
  \end{ocg}%
  \hspace{-\hsize}%
  \begin{ocg}{fig:30b3off}{fig:30b3off}{0}%
  \end{ocg}%
  \begin{ocg}{fig:30b3on}{fig:30b3on}{1}%
    \includegraphics[width=\hsize]{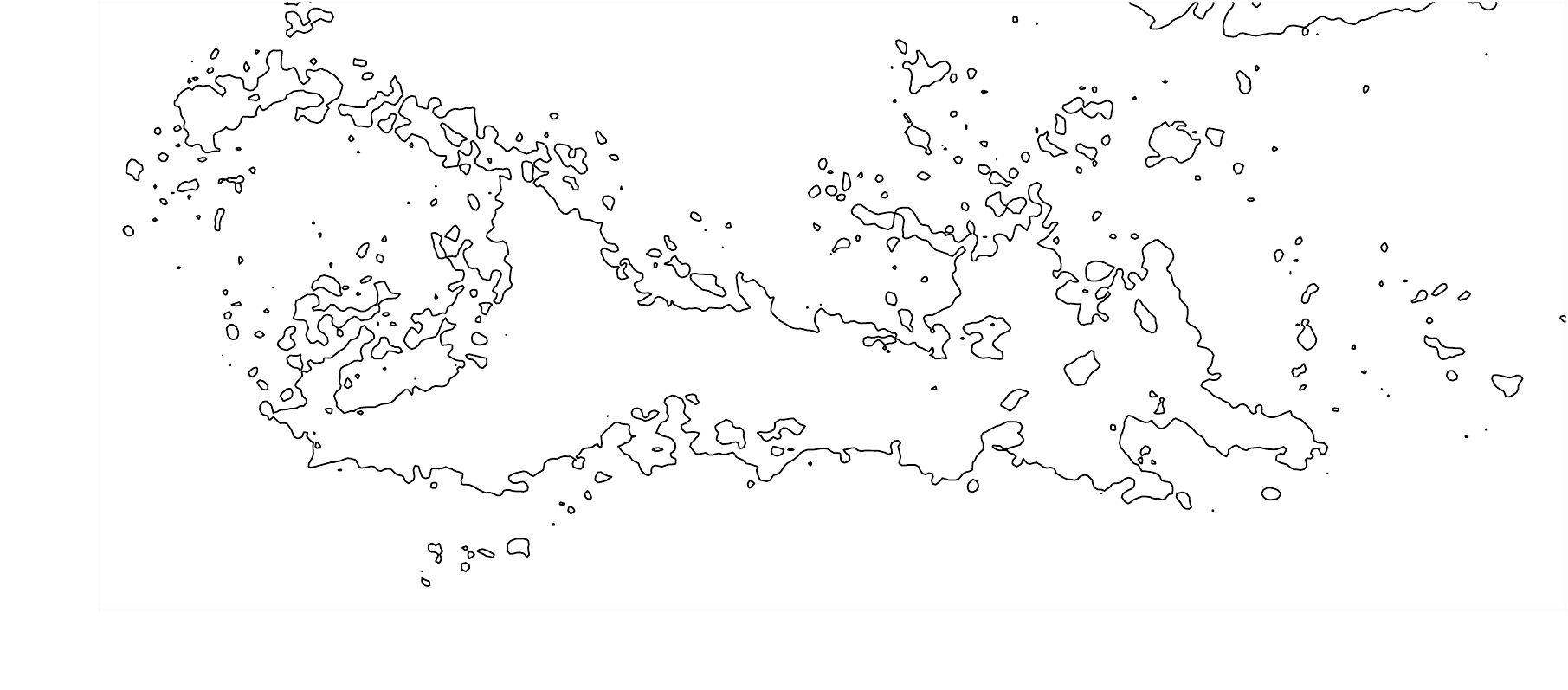}%
  \end{ocg}%
  \hspace{-\hsize}%
  \begin{ocg}{fig:30b4off}{fig:30b4off}{0}%
  \end{ocg}%
  \begin{ocg}{fig:30b4on}{fig:30b4on}{1}%
    \includegraphics[width=\hsize]{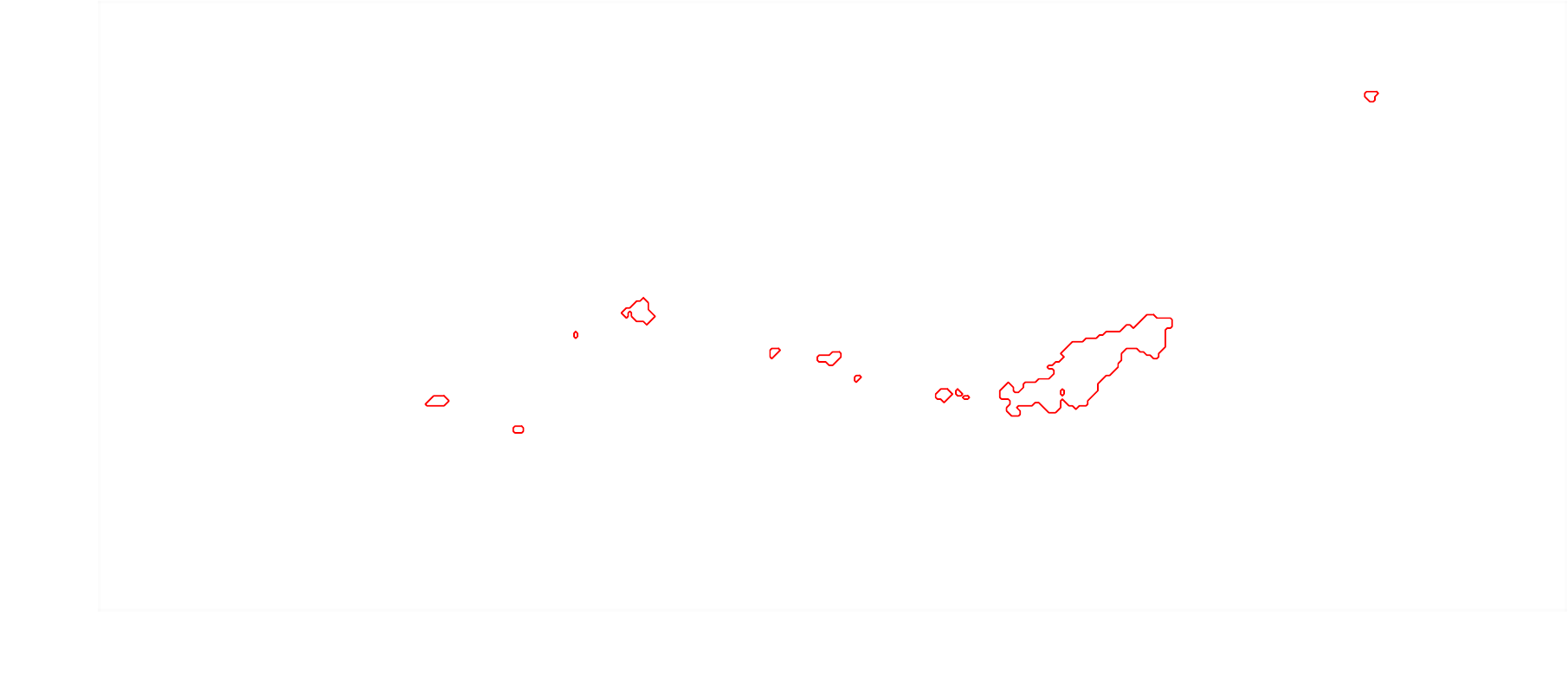}%
  \end{ocg}%
  \hspace{-\hsize}%
  \begin{ocg}{fig:30b5off}{fig:30b5off}{0}%
  \end{ocg}%
  \begin{ocg}{fig:30b5on}{fig:30b5on}{1}%
    \includegraphics[width=\hsize]{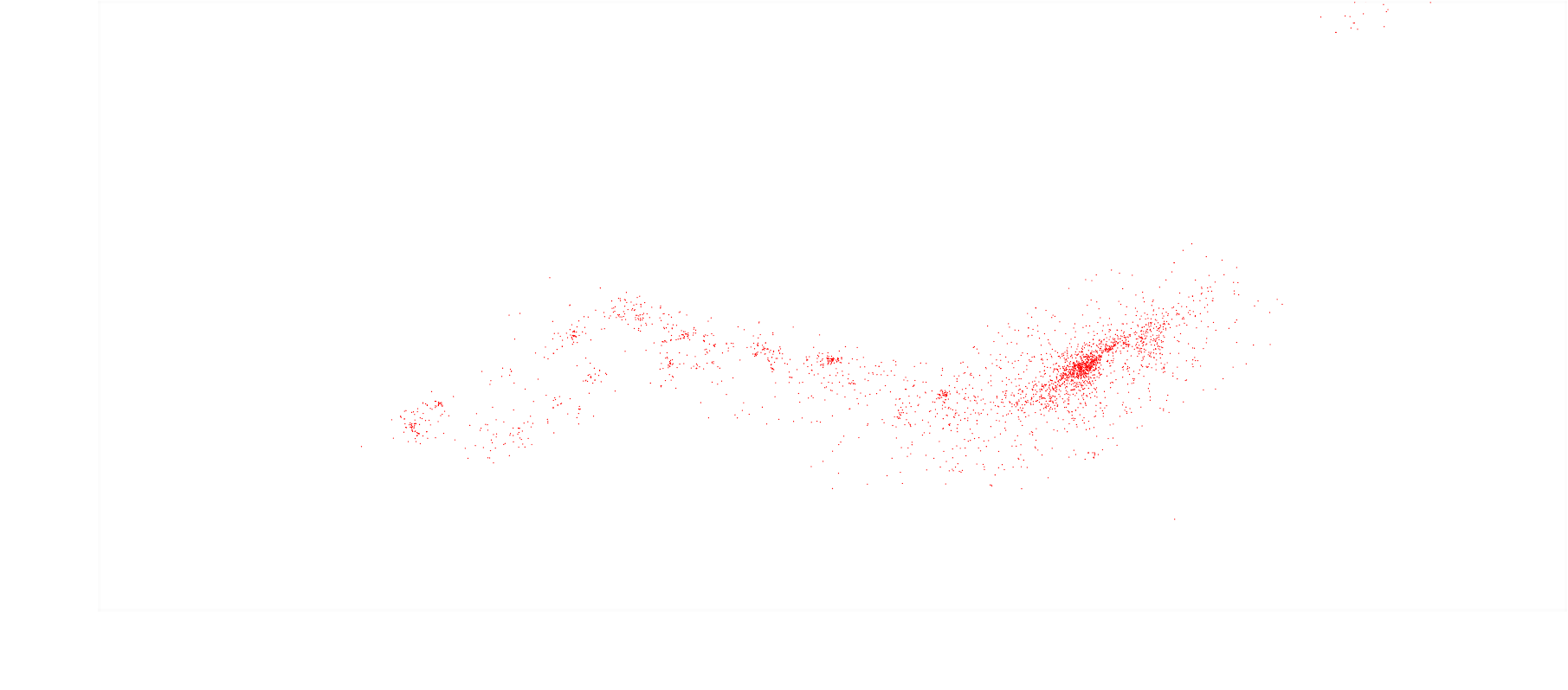}%
  \end{ocg}%
  \hspace{-\hsize}%
  \begin{ocg}{fig:30b6off}{fig:30b6off}{0}%
  \end{ocg}%
  \begin{ocg}{fig:30b6on}{fig:30b6on}{1}%
    \includegraphics[width=\hsize]{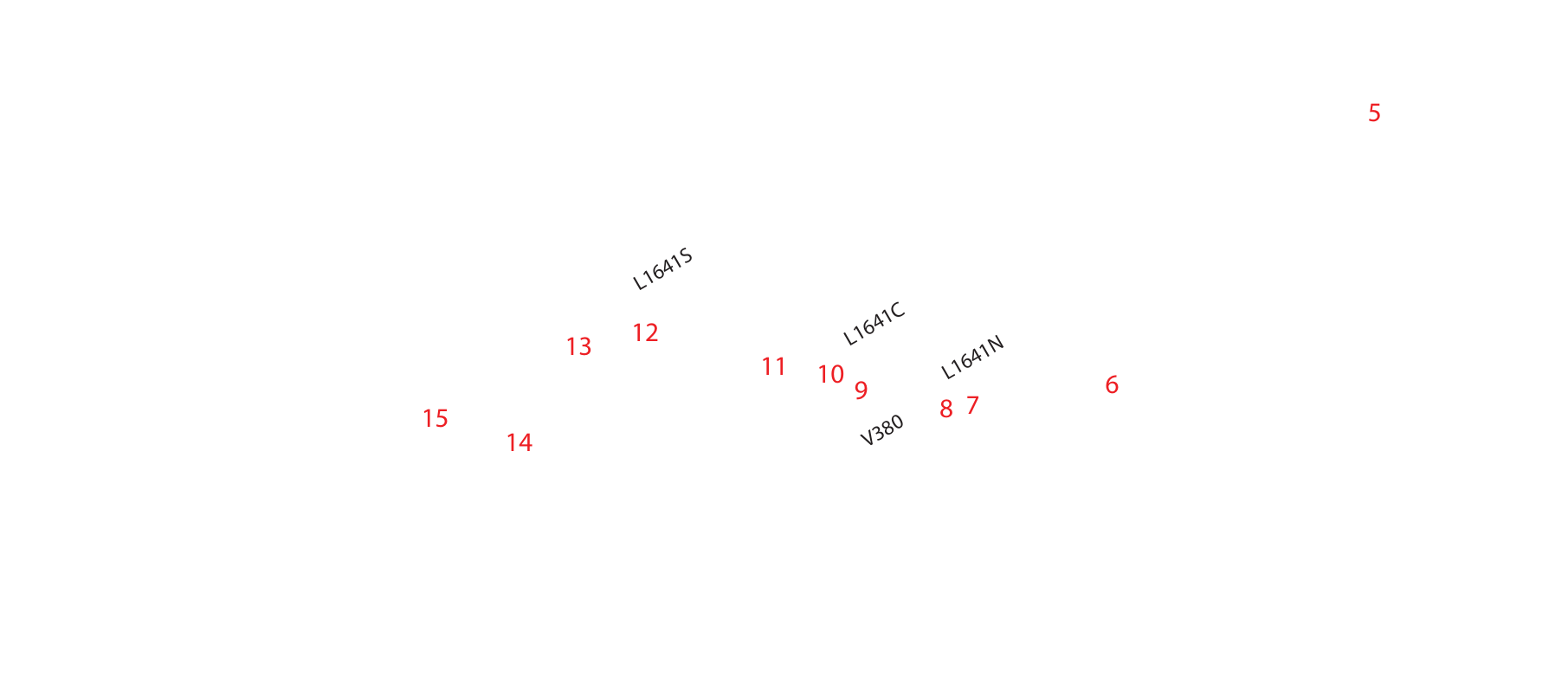}%
  \end{ocg}%
  \caption{%
    The results of the cluster finding algorithm in Orion~A using the
    2MASS Point Source Catalog.  The red contours shows all surface
    density detections $3\sigma$ above the background, while the black
    contour corresponds to $A_K = \SI{0.3}{mag}$.  When displayed in Adobe
    Acrobat, it is possible to hide the
    \ToggleLayer{fig:30b2on,fig:30b2off}{\protect\cdbox{Grid}} lines,
    the black
    \ToggleLayer{fig:30b3on,fig:30b3off}{\protect\cdbox{Extinction}}
    contours, the red contours corresponding to the
    \ToggleLayer{fig:30b4on,fig:30b4off}{\protect\cdbox{Clusters}},
    the red dots representing the
    \ToggleLayer{fig:30b5on,fig:30b5off}{\protect\cdbox{Megeath et
        al.\ YSOs'}}, 
    and the clusters'
    \ToggleLayer{fig:30b6on,fig:30b6off}{\protect\cdbox{Names}}.}
  \label{fig:11}
\end{figure*}

\begin{figure}[t]
  \centering
  \includegraphics[width=\hsize]{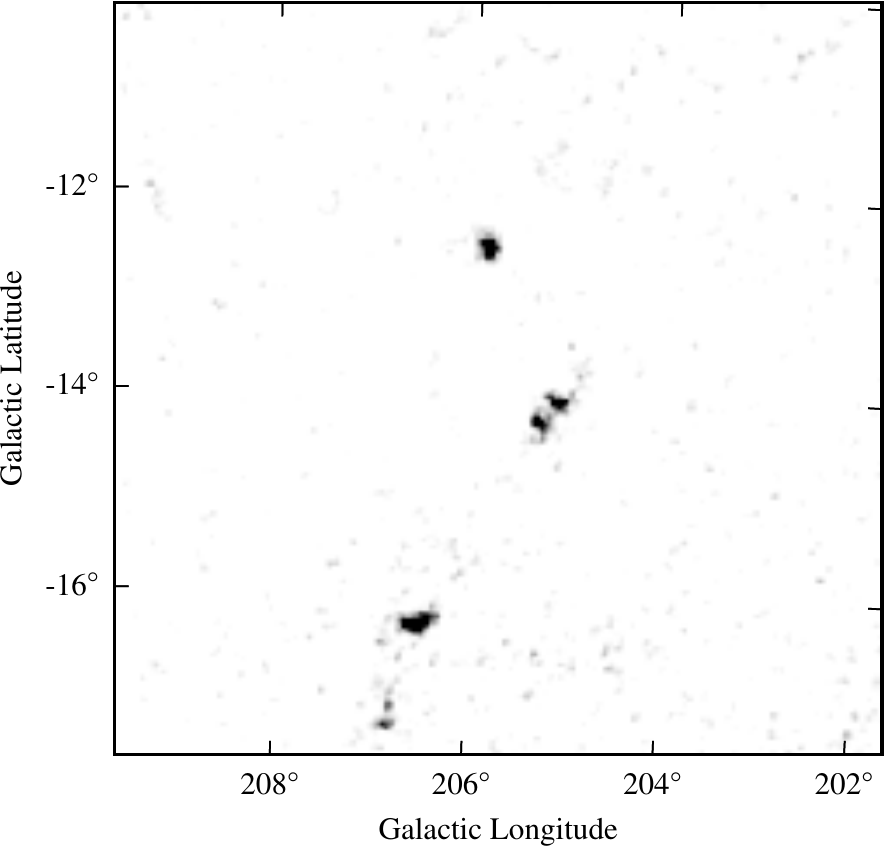}%
  \hspace{-\hsize}%
  \begin{ocg}{fig:30c2off}{fig:30c2off}{0}%
  \end{ocg}%
  \begin{ocg}{fig:30c2on}{fig:30c2on}{1}%
    \includegraphics[width=\hsize]{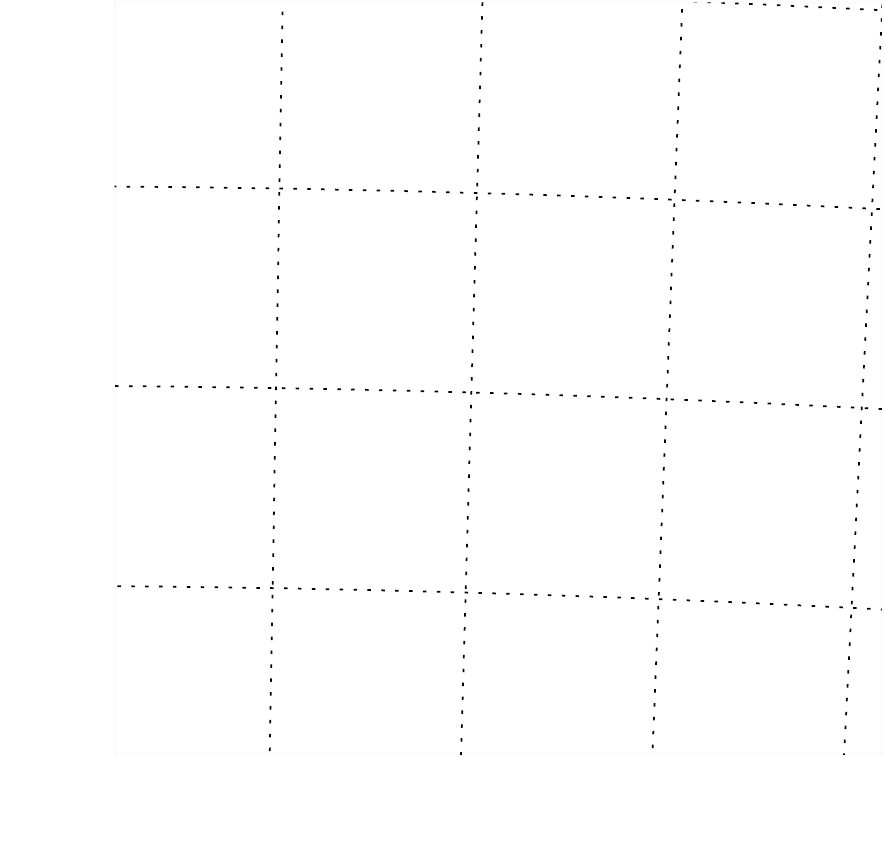}%
  \end{ocg}%
  \hspace{-\hsize}%
  \begin{ocg}{fig:30c3off}{fig:30c3off}{0}%
  \end{ocg}%
  \begin{ocg}{fig:30c3on}{fig:30c3on}{1}%
    \includegraphics[width=\hsize]{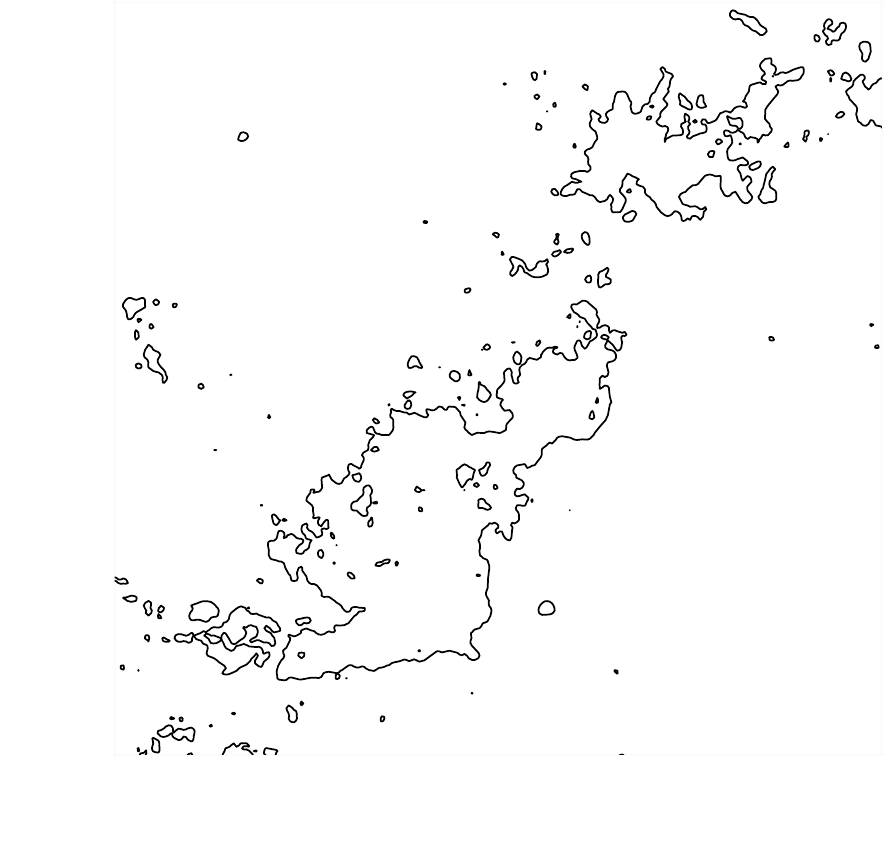}%
  \end{ocg}%
  \hspace{-\hsize}%
  \begin{ocg}{fig:30c4off}{fig:30c4off}{0}%
  \end{ocg}%
  \begin{ocg}{fig:30c4on}{fig:30c4on}{1}%
    \includegraphics[width=\hsize]{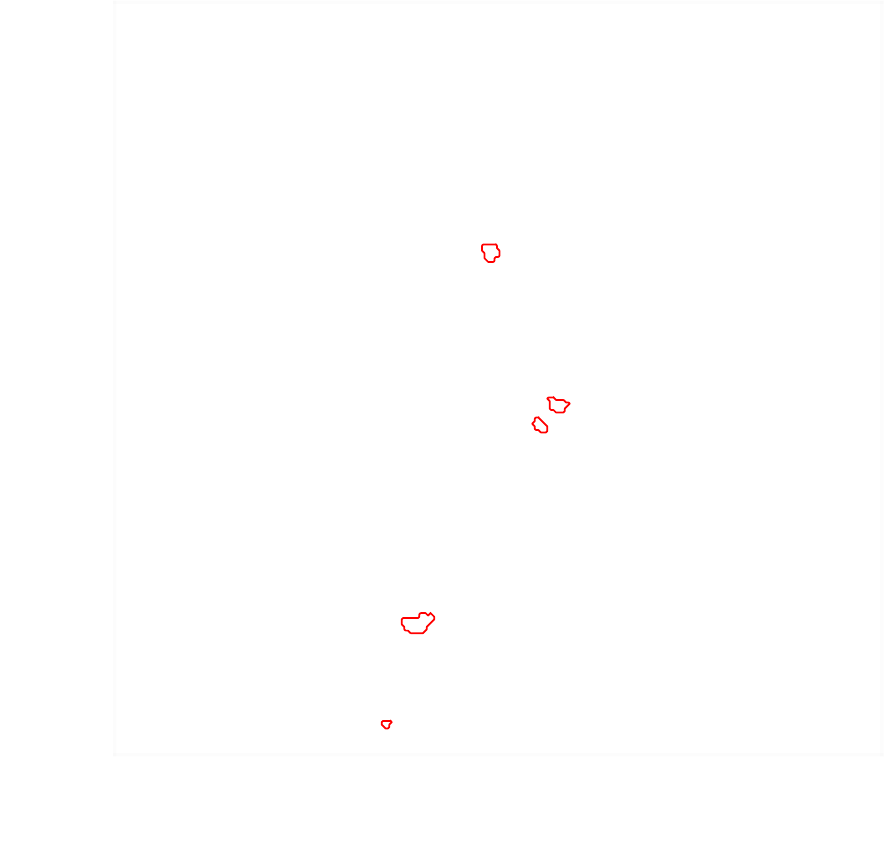}%
  \end{ocg}%
  \hspace{-\hsize}%
  \begin{ocg}{fig:30c5off}{fig:30c5off}{0}%
  \end{ocg}%
  \begin{ocg}{fig:30c5on}{fig:30c5on}{1}%
    \includegraphics[width=\hsize]{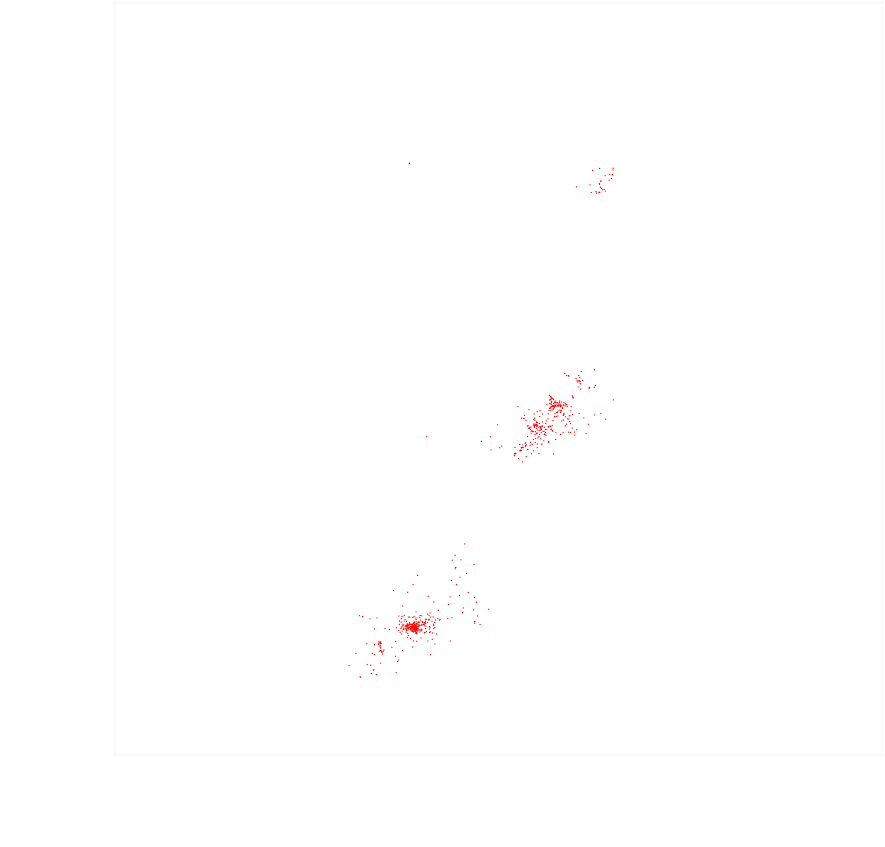}%
  \end{ocg}%
  \hspace{-\hsize}%
  \begin{ocg}{fig:30c6off}{fig:30c6off}{0}%
  \end{ocg}%
  \begin{ocg}{fig:30c6on}{fig:30c6on}{1}%
    \includegraphics[width=\hsize]{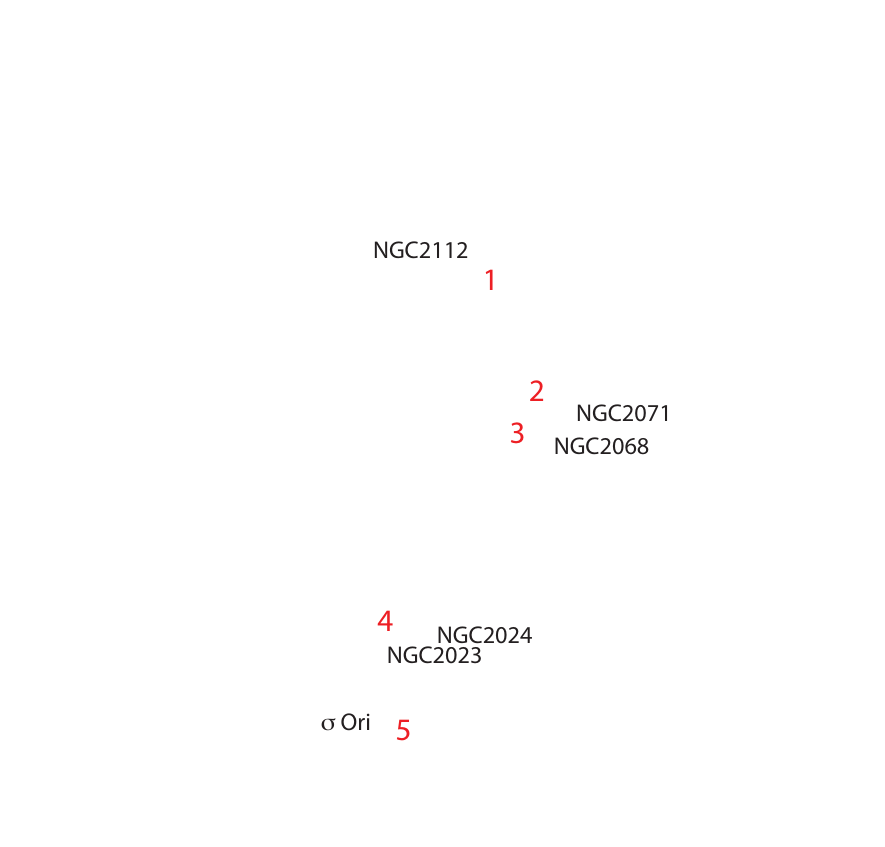}%
  \end{ocg}%
  \caption{%
    The results of the cluster finding algorithm in Orion~B using the
    2MASS Point Source Catalog (see caption of Fig.~\ref{fig:11} for
    the legend).  Note that NGC~2023 is below the detection
    threshold and appears only as a weak smudge in this image. When
    displayed in Adobe Acrobat, it is possible to hide the
    \ToggleLayer{fig:30c2on,fig:30c2off}{\protect\cdbox{Grid}}, the
    \ToggleLayer{fig:30c3on,fig:30c3off}{\protect\cdbox{Extinction}},
    the
    \ToggleLayer{fig:30c4on,fig:30c4off}{\protect\cdbox{Clusters}},
    the red dots representing the
    \ToggleLayer{fig:30b5on,fig:30b5off}{\protect\cdbox{Megeath et
        al.\ YSOs'}}, 
    or the
    \ToggleLayer{fig:30c6on,fig:30c6off}{\protect\cdbox{Names}}.}
  \label{fig:12}
\end{figure}

\begin{figure}[t]
  \centering
  \includegraphics[width=\hsize]{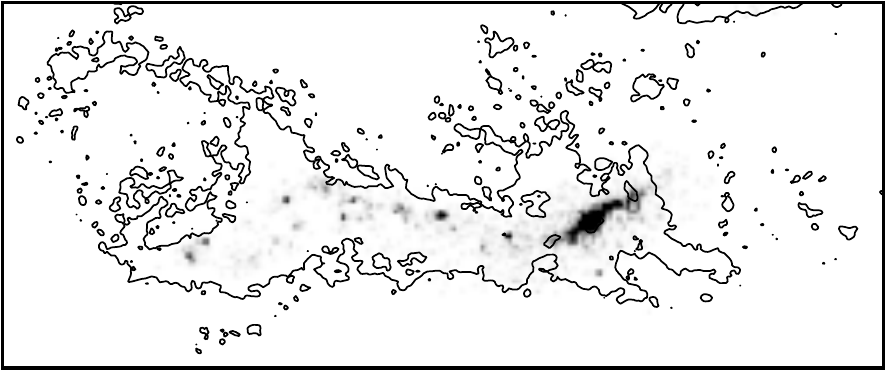}
  \\[1mm]
  \includegraphics[width=\hsize]{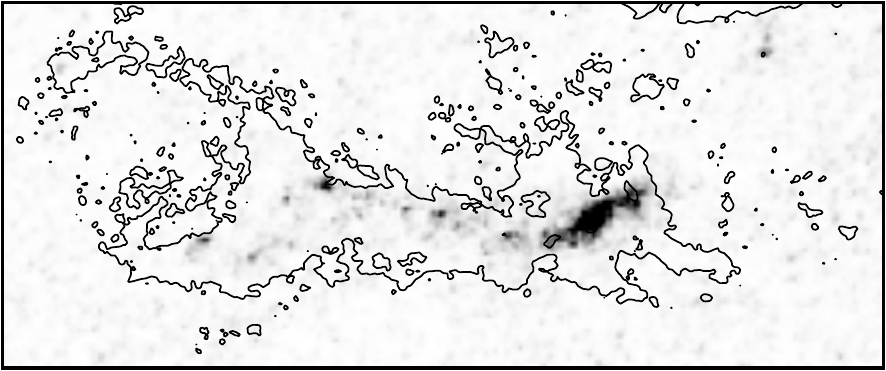}%
  \caption{%
    A Gaussian-kernel smoothed density map of the
    \citet{2016AJ....151....5M} YSO list in Orion~A (top) to
    compare with our density map (bottom).}
  \label{fig:122}
\end{figure}

\begin{figure}[t]
  \centering
  \includegraphics[width=0.49\hsize]{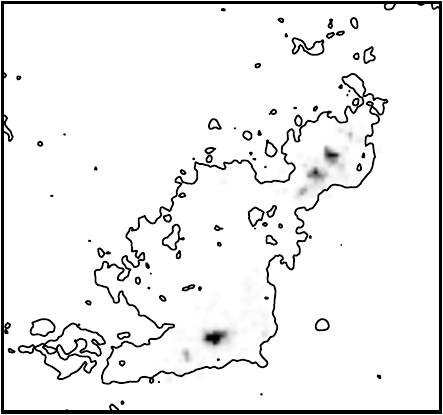}\hfill
  \includegraphics[width=0.49\hsize]{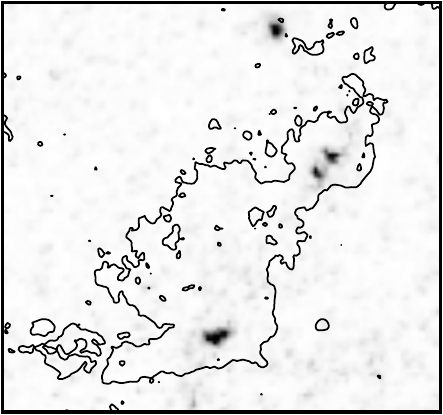}%
  \caption{A Gaussian-kernel smoothed density map of the
    \citet{2016AJ....151....5M} YSO list in Orion~B (left) to
    compare with our density map (right).}
  \label{fig:123}
\end{figure}

\begin{figure}[t]
  \centering
  \includegraphics[width=\hsize]{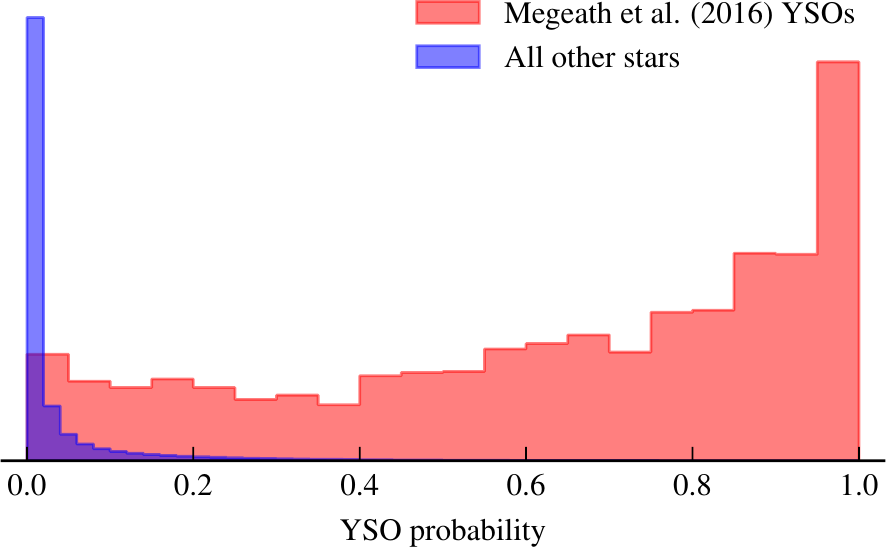}%
  \caption{The distribution of $P_\mathrm{YSO}$, the probability
    that our algorithm assigns to each star to be a YSO, for the
    \citet{2016AJ....151....5M} YSO candidates and for all other
    objects in the 2MASS point source catalog.}
  \label{fig:124}
\end{figure}

{
  \setlength{\tabcolsep}{4pt}%
  \begin{table*}
\caption{The YSO clusters identified in Orion~A and B.\label{tab:3}}
\begin{tabular}{ccccccccccc}
\toprule
ID & $\ell$ & $b$ & R.A. & Dec. & Stars & $R$ & $A_K$ & Cloud & Simbad & Type \\
 & $\mathrm{{}^{\circ}}$ & $\mathrm{{}^{\circ}}$ & $\mathrm{{}^{\circ}}$ & $\mathrm{{}^{\circ}}$ &  & $\mathrm{{}^{\prime}}$ & $\mathrm{mag}$ &  &  &  \\
\midrule
1 & $205.8569$ & $-12.6041$ & $ 88.4398$ & $+00.4283$ & $\phantom{0}\phantom{0}88\pm14\phantom{0}$ & $\phantom{0}5.2$ & $ 0.2$ & Ori B & NGC 2112 & OpC \\
2 & $205.1234$ & $-14.1127$ & $ 86.7727$ & $+00.3486$ & $\phantom{0}\phantom{0}78\pm13\phantom{0}$ & $\phantom{0}4.9$ & $ 1.6$ & Ori B & NGC 2071 & RNe \\
3 & $205.2984$ & $-14.3236$ & $ 86.6667$ & $+00.0995$ & $\phantom{0}\phantom{0}56\pm11\phantom{0}$ & $\phantom{0}4.1$ & $ 1.4$ & Ori B & M 78 & RNe \\
4 & $206.4996$ & $-16.3341$ & $ 85.4332$ & $-01.8659$ & $\phantom{0}188\pm23\phantom{0}$ & $\phantom{0}7.1$ & $ 2.1$ & Ori B & NGC 2024 & Cl* \\
5 & $206.8020$ & $-17.3464$ & $ 84.6710$ & $-02.5928$ & $\phantom{0}\phantom{0}15\pm4\phantom{0}\phantom{0}$ & $\phantom{0}2.5$ & $ 0.1$ & Ori A & sig Ori & MGr \\
6 & $208.9681$ & $-19.3846$ & $ 83.7998$ & $-05.3541$ & $1601\pm49\phantom{0}$ & $19.8$ & $ 1.1$ & Ori A & M 42 & HII \\
7 & $209.9598$ & $-19.6021$ & $ 84.0276$ & $-06.2867$ & $\phantom{0}\phantom{0}\phantom{0}8\pm3\phantom{0}\phantom{0}$ & $\phantom{0}1.9$ & $ 0.6$ & Ori A & --- & --- \\
8 & $210.1018$ & $-19.6036$ & $ 84.0865$ & $-06.4071$ & $\phantom{0}\phantom{0}27\pm6\phantom{0}\phantom{0}$ & $\phantom{0}3.1$ & $ 1.1$ & Ori A & LDN 1641N & Cl* \\
9 & $210.7790$ & $-19.4832$ & $ 84.4813$ & $-06.9247$ & $\phantom{0}\phantom{0}\phantom{0}5\pm3\phantom{0}\phantom{0}$ & $\phantom{0}1.5$ & $ 0.6$ & Ori A & [BDB2003] G210.80-19.50 & Cl* \\
10 & $210.9941$ & $-19.3347$ & $ 84.7061$ & $-07.0406$ & $\phantom{0}\phantom{0}35\pm9\phantom{0}\phantom{0}$ & $\phantom{0}3.6$ & $ 1.7$ & Ori A & LDN 1641C & Cl* \\
11 & $211.4326$ & $-19.2912$ & $ 84.9301$ & $-07.3924$ & $\phantom{0}\phantom{0}12\pm5\phantom{0}\phantom{0}$ & $\phantom{0}2.1$ & $ 2.2$ & Ori A & --- & --- \\
12 & $212.4806$ & $-18.9864$ & $ 85.6458$ & $-08.1475$ & $\phantom{0}\phantom{0}91\pm13\phantom{0}$ & $\phantom{0}5.3$ & $ 0.6$ & Ori A & 2MASS J05422695-0809173 & Y*O \\
13 & $212.9849$ & $-19.1519$ & $ 85.7057$ & $-08.6486$ & $\phantom{0}\phantom{0}\phantom{0}4\pm2\phantom{0}\phantom{0}$ & $\phantom{0}1.2$ & $ 1.4$ & Ori A & [BDB2003] G212.98-19.15 & Cl* \\
14 & $213.4450$ & $-19.8466$ & $ 85.2625$ & $-09.3407$ & $\phantom{0}\phantom{0}10\pm4\phantom{0}\phantom{0}$ & $\phantom{0}2.1$ & $ 0.6$ & Ori A & --- & --- \\
15 & $214.0756$ & $-19.6278$ & $ 85.7219$ & $-09.7820$ & $\phantom{0}\phantom{0}32\pm8\phantom{0}\phantom{0}$ & $\phantom{0}3.4$ & $ 1.1$ & Ori A & [BDB2003] G214.06-19.62 & Cl* \\
\bottomrule
\end{tabular}
\end{table*}

}

{
  \setlength{\tabcolsep}{4pt}%
  \begin{table}
\caption{Correspondence between the clusters identified in this work, marked 
with the subscript 1, and the ones identified from a smoothed version of the YSO catalog 
\citet{2016AJ....151....5M}.\label{tab:4}}
\begin{tabular}{cccccc}
\toprule
$\ell$ & $b$ & Stars${}_1$ & Stars${}_2$ & $R$ & Cloud \\
$\mathrm{{}^{\circ}}$ & $\mathrm{{}^{\circ}}$ &  &  & $\mathrm{{}^{\prime}}$ &  \\
\midrule
$214.0472$ & $-19.6402$ & $\phantom{0}\phantom{0}12\pm4\phantom{0}\phantom{0}$ & $\phantom{0}8.9$ & $\phantom{0}1.9$ & Ori A \\
$212.9892$ & $-19.1435$ & $\phantom{0}\phantom{0}10\pm5\phantom{0}\phantom{0}$ & $11.5$ & $\phantom{0}2.1$ & Ori A \\
$212.4676$ & $-19.0192$ & $\phantom{0}\phantom{0}\phantom{0}8\pm3\phantom{0}\phantom{0}$ & $\phantom{0}3.4$ & $\phantom{0}1.2$ & Ori A \\
$210.9717$ & $-19.3367$ & $\phantom{0}\phantom{0}28\pm8\phantom{0}\phantom{0}$ & $39.5$ & $\phantom{0}3.3$ & Ori A \\
$210.0881$ & $-19.5955$ & $\phantom{0}\phantom{0}23\pm6\phantom{0}\phantom{0}$ & $31.1$ & $\phantom{0}2.8$ & Ori A \\
$209.5134$ & $-19.6566$ & $\phantom{0}\phantom{0}\phantom{0}4\pm2\phantom{0}\phantom{0}$ & $\phantom{0}3.4$ & $\phantom{0}1.2$ & Ori A \\
$209.0159$ & $-19.3924$ & $\phantom{0}929\pm36\phantom{0}$ & $873.3$ & $12.9$ & Ori A \\
$208.5245$ & $-19.1518$ & $\phantom{0}\phantom{0}68\pm11\phantom{0}$ & $51.6$ & $\phantom{0}4.3$ & Ori A \\
$208.4171$ & $-19.1892$ & $\phantom{0}\phantom{0}20\pm5\phantom{0}\phantom{0}$ & $12.1$ & $\phantom{0}2.2$ & Ori A \\
$208.3900$ & $-19.0734$ & $\phantom{0}\phantom{0}15\pm4\phantom{0}\phantom{0}$ & $10.3$ & $\phantom{0}2.1$ & Ori A \\
$206.5222$ & $-16.3557$ & $\phantom{0}135\pm19\phantom{0}$ & $158.5$ & $\phantom{0}5.5$ & Ori B \\
$205.3203$ & $-14.3258$ & $\phantom{0}\phantom{0}39\pm9\phantom{0}\phantom{0}$ & $28.2$ & $\phantom{0}3.3$ & Ori B \\
$205.1356$ & $-14.0971$ & $\phantom{0}\phantom{0}61\pm12\phantom{0}$ & $55.6$ & $\phantom{0}4.2$ & Ori B \\
\bottomrule
\end{tabular}
\end{table}

}

We have applied the method proposed in this paper to 2MASS data of the
Orion~A and B molecular cloud complexes.  The regions selected ensure
that we can verify the reliability of the algorithm proposed here using
some of the best studied objects in the sky.  In particular, the
populations of embedded clusters for both clouds have been the
targets of extensive observational campaigns using ground-based, 
near-infrared \citep{1991ApJ...371..171L}) and millimeter-wave
\citep{2016arXiv160906366L} surveys as well as space-based, mid-infrared
\textit{Spitzer Space Telescope} \citep{2012AJ....144..192M,
2016AJ....151....5M}, and \textit{Chandra} X-ray 
\citep{2008ApJ...677..401P} surveys.  Additionally, the distance
of these clouds ensures that the 2MASS $H$-band data are, in absence of
extinction, complete for YSOs: that is, the cluster HLF at the distance of
Orion is essentially entirely within the 2MASS $H$-band limiting
magnitude ($\sim \SI{15}{mag}$).

\begin{figure}[th]
  \centering
  \includegraphics[width=\hsize]{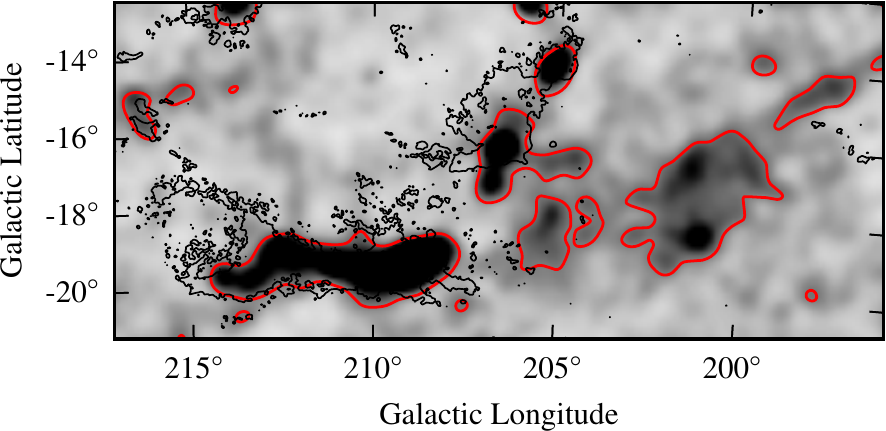}
  \caption{A low-resolution version of the YSO density in Orion.
    The grey area at $\ell \simeq 205^\circ$ is the Orion OB Ib
    association, while the lighter area to the right (around $\ell
    \simeq 201^\circ$) is OB Ia, containing the 25~Ori cluster (the
    grey spot at $\ell \simeq 201^\circ$ and $b \simeq -18^\circ$).}
  \label{fig:13}
\end{figure}

Figures~\ref{fig:11} and \ref{fig:12} show the density of fiducial
YSOs measured in Orion~A and B.  These maps have been produced by our
pipeline together with \textsc{Nicer} \citep{2001A&A...377.1023L} and
\textsc{Nicest} \citep{2009A&A...493..735L} extinction maps from the
2MASS Point Source Catalogue (see \citealp{2011A&A...535A..16L} for
details on the data selection and extinction map construction).  The
algorithm has been run in conditions similar to the simulations
described above: that is, we have used two different stellar
populations, one associated with the background and characterized by
exponential number counts, and one associated with the YSOs and
characterized by Gaussian number counts (with parameters consistent
with the $H$-band luminosity function of
\citealp{2002ApJ...573..366M}).  Using an angularly close control
field we also measured the distribution of intrinsic colors of stars
and the shape of the completeness function: the latter has been
modeled using an complementary error function erfc, as described in
Appendix~\ref{sec:k-band-probability}. This choice makes it possible
to use the entire 2MASS catalogue without any magnitude cut (which
would increase the noise in the final data products).  The maps
produced have a pixel size of \SI{1.5}{arcmin} and a weight function $W
\propto \omega$ in turn proportional to a Gaussian with $\mathit{FWHM}
= \SI{3}{arcmin}$.  We have used a relatively large beam in these maps
to increase the sensitivity of our algorithm and to minimize the
effects of the biases shown in the simulations described in
Sect.~\ref{sec:simulations}, while still retaining in most situations
the ability to distinguish different clusters (i.e., avoid confusion
effects at the distance of Orion).

Since we have at our disposal the covariance map of these
measurements, we have assessed the reliability of each density peak in
these figures.  The red contours in the figures show the areas (larger
than 2 pixels) where the local YSO density exceeds
\SI{1.5}{stars/pixel}, corresponding approximately to a
signal-to-noise ratio larger than 3.5: note how some regions in black
in Fig.~\ref{fig:11} do not reach the threshold because of the large
error associated with them (mostly due to the high extinction values
there).

Table~\ref{tab:3} shows the YSO clusters identified in the Orion~A and
B areas using our simple prescription, together with the most relevant
parameters. In some cases we clearly see that angularly close clusters
appear as a single contour in our maps: the simple procedure used here
to define clusters, the relatively coarse resolution used, and the
cluster morphology itself prevent us from deblending some close
objects. An extreme case of this situation might be the ISF (the
Integral Shaped Filament) cluster, where the limitations due to
angular resolution would make it difficult to resolve and separate
smaller clusters if they exist in such a very populous region. We note
that the ISF cluster encompasses M42, the Trapezium and ONC clusters
as well as an extended population of YSOs along the ISF.

The radius $R$ reported in the table corresponds to the radius of a
circle that would occupy the same area as the identified cluster,
i.e.\ to the connected region of the sky where the inferred
density of YSOs exceeds the background by $3\sigma$.  At the estimated
distance of Orion, \SI{413}{pc} \citep{2009ApJ...700..137R}, $1'$
corresponds to \SI{0.12}{pc}: therefore, the clusters identified have
radii spanning from $\sim \SI{2.4}{pc}$ to $\sim \SI{0.15}{pc}$.

The well known clusters in these clouds are correctly identified by
our procedure.  It is interesting to compare Table~\ref{tab:3} with
clusters identified independently using much more secure data.  Among
the ones at our disposal, the recent catalog of embedded YSOs
obtained by \citet{2016AJ....151....5M} using the \textit{Spitzer
  Space Telescope} and the \textit{Chandra} observatory is probably
the most secure and complete: we will therefore focus on this catalog.
Since our definition of a cluster is based on an somewhat arbitrary
parameters (signal-to-noise threshold, minimum number of pixels, no
correction for the stars missed at the boundaries), and since
different, more-or-less, arbitrary parameters are also used by
\citet{2016AJ....151....5M}, we find it more appropriate and fair to
make a comparison after we first homogenize the data. Specifically, we
take \citet{2016AJ....151....5M} YSO list and we make out of it
a density map using a Gaussian kernel of the same size of the one used
for our map. Figures \ref{fig:122} and \ref{fig:123} show the results
obtained for Orion~A and B, which clearly compares very well with our
own maps, derived purely from the 2MASS point source catalog. The most
relevant structures are present in both maps and have very similar
shapes; the only differences are the noise present in our maps
(however at a relatively low level), and the limited coverange of
the \textit{Spitzer} derived density maps.

The qualitative similarity of these maps can be quantified if compare
clusters identified in both maps using the same
criteria. Table~\ref{tab:4} shows a list of clusters identified in the
smoothed \citet{2016AJ....151....5M} maps using a fix density
threshold (set to \SI{1.5}{stars/pixel}). In this table we compare the
number of \textit{Spitzer} YSOs with the number of YSOs predicted from
the integral of the $\sigma_\mathrm{YSOs}$ over the area of each
cluster as defined from Megeath et al.\ density map, together with the
computed $1$-$\sigma$ error. It is clear that in almost all cases we
find an excellent agreement, although in many cases our estimates are
slightly larger than the ones by Megeath et al. We can speculate that
this is due to the presence of class~III YSO, which likely would
be missed by \textit{Spitzer}. Indeed, a comparison of the two panels
of Fig.~\ref{fig:122} shows that the bottom panel, corresponding to
our density map, has spatially more extended clusters than the top
panel, corresponding to Megeath et al.\ density map.

As discussed earlier on, our algorithm is a statistical one and works
best when it is applied to a sizeable number of stars. However, we can
also push it and associate to each single star a probability of being
a YSO: to this purpose, for the $n$-th star we can compute
\begin{equation}
  \label{eq:47}
  P_i = \frac{\sigma_i(\vec x_n) p_i(m_n - A_n)}{\sum_{j=1}^L
    \sigma_j(\vec x_n) p_j(m_n - A_n)} \; .
\end{equation}
Note how this quantity resembles the term within the outer sum of
Eq.~\eqref{eq:30}.  Figure~\ref{fig:124} shows the distribution of
$P_\mathrm{YSO}$ (that is, the distribution in the probabilities
assigned to each object to be a YSO) for the
\citet{2016AJ....151....5M} YSO candidates and for all the other
objects. It is clear how all the other objects have $P_\mathrm{YSO}$
that is concentrated around zero, while the YSO candidates have
a much broader distribution that extends to unity.  For these latter
objects the distribution, in addition to a substal peak at
$P_\mathrm{YSO} = 1$, shows a long tail up to small values of
$P_\mathrm{YSO}$: this is not unexpected, since our identification
is only statistical (and therefore we cannot identify unambigously
YSOs). Note also how the relatively low values of
$P_\mathrm{YSO}$ for some genuine YSOs in our algorithm are
compensated by the small tail in the distribution of field stars (this
works because there are many more field stars than YSOs, a fact
that is accounted for in the algorithm).

\subsection{Sensitivity to the distributed population} 

Recently, \citealp{2016arXiv160904948K}, have identified a rich and
well-defined stellar population of about 2\,000 objects, mostly M
stars without extinction or infrared-excesses, as the low-mass
counterpart to the Orion OB Ib subgroup (the Orion belt
population). This low-mass population is not obviously clustered but
instead appears to be distributed across $\sim 10$ square degrees and
the authors speculate that it could represent the evolved counterpart
of a Orion nebula-like cluster.  While more data is needed to test
this scenario, it is clear that much can be learned about the origin
of stellar OB associations and the dispersal of clusters into the
Galactic field if one is able to trace in a robust manner the
distribution of the slightly older and more expanded populations
surrounding star-forming molecular clouds.

We now investigate the extent to which the technique proposed here is
suitable for detection of looser, more expanded distributions of young
stars, in particular the low-mass counterpart to the Orion OB
association presented in \citealp{2016arXiv160904948K}.  For this
purpose, we have built a lower resolution map of the region, employing
a \textit{FWHM} of \SI{30}{arcmin}.

Figure~\ref{fig:13} shows that, surprisingly, we are well able to
recover the stellar over-density of the Ori~Ib population, and for the
first time, the stellar over-density of the Ori~Ia population. These
over-densities are likely to be created by low-mass stars as 2MASS is
still sensitive to the peak of the IMF for the putative distance and
age of these subgroups. An analysis of the substructure seen in the
distributed population visible in Figure~\ref{fig:13} above the noise
pattern is beyond the scope of this paper, but will best addressed
once Gaia parallaxes are generally available. Of relevance for this
paper is that the ability of the method to trace the dispersed
population from existing all sky data opens a new window on the
unsolved problem of the origins of OB association and cluster
dispersal in to the field.



\section{Conclusions}
\label{sec:conclusions}

The following items summarize the main results presented in this
paper:
\begin{itemize}
\item We have developed a new method to discover and characterize
  deeply embedded star clusters.
\item The method is able to statistically classify objects as field
  stars or YSOs and corrects for the effects of differential
  extinction.
\item We have provided expressions for the covariance of the inferred
  densities and we have validated both the method and the analytic
  expression for the covariance with a set of simple but realistic
  simulations.
\item We have applied the new method to 2MASS point sources observed
  in the Orion~A and B and we have shown that we can identify and
  characterize well protostellar clusters in these regions, as well as
  detect much looser associations such as the OB 1a and 1b subgroups.
\end{itemize}

Finally, we note that the method proposed here can be easily extended
to multi-band observations by using suitable probability distributions
$p_i(\vec m)$ for various populations as a function of the magnitude
vector $\vec m$.  Its implementation would be essentially identical,
with minor modifications to take into account the different effects of
extinction on different bands.  The natural use of such an extension
would be in the context of techniques such as the one proposed by
\citet{2016A&A...585A..78J} which are able to recover the extinction
from a complex analysis of multi-band data.

\begin{acknowledgements}
  This research has made use of the 2MASS archive, provided by
  NASA/IPAC Infrared Science Archive, which is operated by the Jet
  Propulsion Laboratory, California Institute of Technology, under
  contract with the National Aeronautics and Space Administration.e
  Additionally, this research has made use of the SIMBAD database,
  operated at CDS, Strasbourg, France.
\end{acknowledgements}

\appendix

\section{Implementation}
\label{sec:implementation-1}

In this section we consider a few details and analytical expressions
useful to evaluate some of the expressions needed to implement the
method proposed here.

\subsection{Algorithm}
\label{sec:algorithm}

The algorithm proposed in this paper is essentially an iterative
application of Eq.~\eqref{eq:30} to the set of densities $\{ \sigma_i
(\vec x) \}$. A few notes, however, are necessary to better implement
the algorithm:
\begin{itemize}
\item The method assumes that the extinction values in the $H$ band are
  available for each star used. In our implementation these values are
  obtained through the use of the \textsc{Nicer} algorithm. Other
  techniques can be used, as long as the extinction is refereed to the
  single star and is \textit{not} an average extinction at the
  location of the star. In our implementation, this requirement limits
  the method to stars for which, in addition to the $H$-band
  measurements, have at least another band photometry.
\item Since the various densities are spatially variable, one needs to
  repeat the iteration of Eq.~\eqref{eq:30} to each point in the map.
\item Equation~\eqref{eq:30} itself needs to be iterated a few times
  (typically around ten, in some cases a little more) before reaching
  a good convergence.  This generally suggests the use of simple
  analytical models for the various functions involved in this
  equation (the probability distributions $p_i(m)$ and the
  completeness function $c(m)$, discussed below).
\end{itemize}

\subsection{{\it H}-band probability distributions and completeness function}
\label{sec:k-band-probability}

\begin{figure}
  \centering
  \includegraphics[width=\hsize]{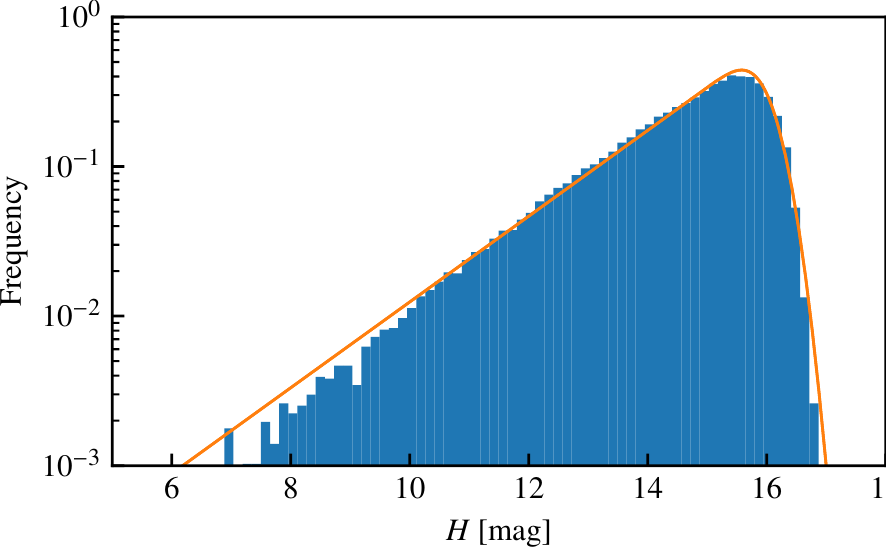}
  \caption{An histogram of $H$ band magnitudes of stars in a control
    field region, together with its best fit (an exponential with a
    complementary error function truncation, see the text).}
  \label{fig:99}
\end{figure}

In this paper we have considered two useful models for the $H$-band
luminosity function: the exponential distribution and the normal
(Gaussian) one.

We parametrize the exponential distribution as
\begin{equation}
  \label{eq:36}
  p(m) = k \beta \e^{\beta m} \propto 10^{\alpha m} \; ,
\end{equation}
where $k$ is a normalization constant that will be found lather on and
$\beta = \alpha \ln 10$.

The normal distribution is parametrized as
\begin{equation}
  \label{eq:40}
  p(m) = \frac{k}{\sqrt{2 \pi s^2}} \e^{-(m - m_0)^2 / 2 s^2} \; ,
\end{equation}
where again $k$ is a normalization constant to be investigated later.

Finally, we model the completeness function $c(m)$ in terms of the
complementary error function $\erfc$. This can justified from an
empirical point of view since the error function has a shape that
resemble the completeness function.  It is also reasonable from a
statistical point of view if the photometric errors are Gaussian: in
this case, one can suppose that the probability that an object be
detected is an integral over a relevant part of the
Gaussian. Specifically, we assume for $c(m)$ the following functional
form:
\begin{equation}
  \label{eq:35}
  c(m) = \frac{1}{2} \erfc \left( \frac{m -
      m_\mathrm{c}}{\sqrt{2 \sigma^2_\mathrm{c}}} \right) \; ,
\end{equation}
where $m_\mathrm{c}$ is the 50\% completeness limit and
$\sigma_\mathrm{c}$ sets the width of the completeness
function. Figure~\ref{fig:99} shows that our simple model for the
control field HLF, the product of $p(m) c(m)$ with $p(m)$ following an
exponential distribution and $c(m)$ described in terms of an erfc,
reproduces well the data.

\subsection{Statistical errors and completeness function}
\label{sec:photometric-errors}

As noted above, statistical errors can be included in our algorithm by
convolving the probability distributions $p_i$ with appropriate
kernels.  We expect to have two main sources for statistical errors:
photometric errors in the $H$-band magnitude measurements and errors
in the extinction measurements.  While the formers are relatively
simple to characterize (typically, they will be provided in the star
catalogue), the latter are a combination of different sources of
errors: the scatter in the intrinsic color of sources and the
photometric errors in each band used.  If extinctions are derived
using the \textsc{Nicer} algorithm, as assumed here, then the variance
associated to each extinction measurement can be computed from the
expression
\begin{equation}
  \label{eq:38}
  \Var(A) \equiv \sigma^2_A = \frac{1}{\vec k^T C^{-1} \vec k} \; ,
\end{equation}
where, following the notation of \citet{2001A&A...377.1023L}, we have
called $\vec k$ is the reddening vector and $C$ the combined
covariance matrix of the observed colors (including both the intrinsic
scatter in the color of unextinguished stars and the individual
photometric errors).

Looking again at Eq.~\eqref{eq:30}, i.e.\ the main equation
representing our method, we see that in reality the $H$-band
magnitudes and the associated extinction measurements typically enter
the problem through the combination $m - A$; moreover, while in the
numerator the combination is $m_n - A_n$ and involves therefore
measurements of both $m$ and $A$ for individual stars, in the
denominator the combination is $m - A_n$, with $m$ integration
variable. Therefore, in presence of errors these two parts of the
expression must be computed separately.

Let us consider first the denominator. Since there the only argument
of $p_i$ with associated errors is $A_n$, the measured extinction for
the $n$-th star, we just need to convolve the magnitude distribution
with a Gaussian kernel with variance provided by Eq.~\eqref{eq:38}. 

In the case of the numerator the situation is slightly more
complicated, because there we find the combination $m_n - A_n$.  Since
$A_n$ is computed from the observed magnitude of each star (including
of course the $H$-band magnitude), $m_n$ and $A_n$ are correlated.  A
simple calculation shows that the variance associated to the
combination $m_n - A_n$ is
\begin{equation}
  \label{eq:39}
  \Var(m_n - A_n) = k_H^2 \sigma^2_A + \sigma^2_H \bigl[ 1 - 2 \,
  (b_{J-H} - b_{H-K}) \, k_H \bigr] \; .
\end{equation}
In this expression the last term is related to the correlation between
$A$ and $m$ and is equal to $\sigma^2_H$, the square of the
photometric error on the $H$-band magnitude. The factors $b_{J-H}$ and
$b_{H-K}$ are quantities that can be computed from Eq.~(12) of
\citet{2001A&A...377.1023L}, while $k_H$ is the reddening vector for
the $H$ band, i.e.\ the ratio $A_H / A$.

\subsection{Convolution integrals}
\label{sec:conv-integr}

The convolution of the luminosity functions \eqref{eq:36} and
\eqref{eq:40} with Gaussian kernels representing the statistical
errors take simple forms.  Calling $\sigma^2_p$ the variance of the
convolution kernel, given depending on the term by Eq.~\eqref{eq:38}
or by Eq.~\eqref{eq:39}, and using the tilde to represent the
convolved $H$-band luminosity functions, we find for the exponential
distribution
\begin{equation}
  \label{eq:41}
  \tilde p(m) = k \beta \e^{\beta m + \sigma^2_p \beta^2 / 2} \; .
\end{equation}
Similarly, for the Gaussian model we find
\begin{equation}
  \label{eq:42}
  \tilde p(m) = \frac{k}{\sqrt{2 \pi (s^2+\sigma^2_p)}} \e^{-(m -
    m_0)^2 / 2 (s^2 + \sigma^2_p)} \; , 
\end{equation}

\subsection{Completeness integrals}
\label{sec:anlyt-expr-relev}

Finally, we need to compute the integral in the denominator of
Eq.~\eqref{eq:30}, involving the $H$-band luminosity function $p(m)$
and the completeness function $c(m)$. A simple change of variable
casts this integral (similar to a convolution) into a more convenient
form:
\begin{equation}
  \label{eq:43}
  \int c(m) p(m-A) \, \diff m = \int c(m+A) p(m) \, \diff m \equiv
  K(A) \; .
\end{equation}
In this equation we have called $K(A)$ the result of the integral,
retaining its explicit dependence on $A$. We now consider the
derivative $K'$ of $K$:
\begin{equation}
  \label{eq:44}
  K'(A) = \int c'(m+A) p(m) \, \diff m \; ,
\end{equation}
and note that, since $c$ is the error function, $c'$ is a Gaussian
function. Since Eq.~\eqref{eq:44} is also (essentially) a convolution,
this allows us to compute $K'(A)$ using formulae similar to the ones
of the previous section: as a result, we see that $K'(A)$ remains
either an exponential or a Gaussian, depending on the function form of
$p(m)$.  

Finally, we need to integrate back $K'(A)$ to obtain $K$. The result
we obtain from this procedure is, in case of an exponential
distribution,
\begin{equation}
  \label{eq:37}
  K(A) = k \e^{\beta (m_\mathrm{c} - A) + \sigma^2_\mathrm{c} \beta^2 / 2}
  \; .
\end{equation}
For the Gaussian distribution the result contains again the error
function:
\begin{equation}
  \label{eq:45}
  K(A) = \frac{k}{2} \erfc \left( \frac{ m_0 - m_\mathrm{c} +
      A}{\sqrt{2(\sigma_\mathrm{c}^2 + s^2)}} \right) \; .
\end{equation}
If statistical errors have to be taken into account, these expressions
simply change into
\begin{equation}
  \label{eq:46}
  K(A) = k \e^{\beta (m_\mathrm{c} - A) + (\sigma^2_\mathrm{c} + \sigma^2_p) \beta^2 / 2}
  \; ,
\end{equation}  
and
\begin{equation}
  \label{eq:459}
  K(A) = \frac{k}{2} \erfc \left( \frac{ m_0 - m_\mathrm{c} +
      A}{\sqrt{2(\sigma_\mathrm{c}^2 + s^2 + \smash[b]{\sigma^2_p})}} \right) \; .
\end{equation}

\bibliographystyle{aa} 
\bibliography{dark-refs}

\end{document}